\documentclass[12pt]{article}
\usepackage{graphicx}
\usepackage{float}
\usepackage{makecell} 
\usepackage{pbox} 
\usepackage{subcaption}
\usepackage[dvipsnames]{xcolor}
\usepackage{stmaryrd}
\usepackage[skip=10pt plus1pt, indent=20pt]{parskip}
\usepackage{amsmath}
\usepackage{amssymb}
\usepackage{indentfirst} 
\usepackage{geometry}
\usepackage[utf8]{inputenc}
\usepackage{lscape} 
\usepackage{float} 
\usepackage{hyperref}
\DeclareMathOperator*{\minimise}{minimise}
\usepackage[backend=bibtex,style=numeric,natbib=true]{biblatex} 
\usepackage{tabularx,ragged2e,booktabs}
\usepackage{authblk}
\usepackage{tcolorbox} 
\usepackage{algorithm} 
\usepackage[noend]{algpseudocode}

\begin{document}

\title{Optimal interruption of \textit{P. vivax} malaria transmission using mass drug administration}

\author[1,2]{Md Nurul Anwar}
\author[1,3,4]{Roslyn I. Hickson}
\author[1]{Somya Mehra}
\author[5,6]{David J. Price}
\author[1,6]{James M. McCaw}
\author[7]{Mark B. Flegg}
\author[1]{Jennifer A. Flegg}
\affil[1]{School of Mathematics and Statistics, The University of Melbourne, Parkville, Australia}
\affil[2]{Department of Mathematics, Bangabandhu Sheikh Mujibur Rahman Science and Technology University, Gopalganj 8100, Bangladesh}
\affil[3]{Australian Institute of Tropical Health and Medicine, and College of Public Health, Medical \& Veterinary Sciences, James Cook University, Townsville, Australia}
\affil[4]{CSIRO, Townsville, Australia}
\affil[5]{Department of Infectious Diseases, The University of Melbourne, at the Peter Doherty Institute for Infection and Immunity, Melbourne, Australia}
\affil[6]{Centre for Epidemiology and Biostatistics, Melbourne School of Population and Global
Health, The University of Melbourne, Parkville, Australia}
\affil[7]{School of Mathematics, Monash University, Melbourne, Australia}
\date{}                     
\setcounter{Maxaffil}{0}
\renewcommand\Affilfont{\itshape\small}
\maketitle
\begin{abstract}
    \textit{Plasmodium vivax} is the most geographically widespread malaria-causing parasite resulting in significant associated global morbidity and mortality. One of the factors driving this widespread phenomenon is the ability of the parasites to remain dormant in the liver. Known as `hypnozoites', they reside in the liver following an initial exposure, before activating later to cause further infections, referred to as `relapses'. As around 79–96\% of infections are attributed to relapses from activating hypnozoites, we expect it will be highly impactful to apply treatment to target the hypnozoite reservoir (i.e., the collection of dormant parasites) to eliminate \textit{P. vivax}. Treatment with radical cure,  for example tafenoquine or primaquine, to target the hypnozoite reservoir is a potential tool to control and/or eliminate \textit{P. vivax}. We have developed a deterministic multiscale mathematical model as a system of integro-differential equations that captures the complex dynamics of \textit{P. vivax} hypnozoites and the effect of hypnozoite relapse on disease transmission. Here, we use our multiscale model to study the anticipated effect of radical cure treatment administered via a mass drug administration (MDA) program. We implement multiple rounds of MDA with a fixed interval between rounds, starting from different steady-state disease prevalences. We then construct an optimisation model with three different objective functions motivated on a public health basis to obtain the optimal MDA interval. We also incorporate mosquito seasonality in our model to study its effect on the optimal treatment regime. We find that the effect of MDA interventions is temporary and depends on the pre-intervention disease prevalence (and choice of model parameters) as well as the number of MDA rounds under consideration. The optimal interval between MDA rounds also depends on the objective (combinations of expected intervention outcomes). We find radical cure alone may not be enough to lead to \textit{P. vivax} elimination under our mathematical model (and choice of model parameters) since the prevalence of infection eventually returns to pre-MDA levels.
\end{abstract}
\section{Introduction}\label{Intro}

\textit{Plasmodium vivax} is a parasite that causes malaria, responsible for 4.5 million cases in 2020 \cite{world2021world}. After an infective mosquito bite, the \textit{P. vivax} parasite triggers a primary infection and can remain dormant (known as a `hypnozoite') within the human liver for a prolonged period before causing a secondary infection known as a `relapse' \cite{white2014modelling,aguas2012modeling}. Because of the relapse characteristics, \textit{P. vivax} has become the most globally widespread parasite and is responsible for significant morbidity and mortality  \cite{antinori2012biology,battle2019mapping}. The reason for hypnozoite activation is still not clear, and the number of hypnozoites established per infective mosquito bite and the recurrence time vary geographically \cite{price2020plasmodium}.\par

When it comes to \textit{P. vivax} for control or elimination, the biological characteristics of \textit{P. vivax} make it more challenging than other malaria parasites because \textit{P. vivax} transmission can be re-established from hypnozoite activation  \cite{mehra2022hypnozoite,price2020plasmodium}. An estimated 79–96\% of the total \textit{vivax} cases are due to hypnozoite activation \cite{robinson2015strategies,huber2021radical,adekunle2015modeling,commons2020estimating}. Thus, targeting the hypnozoite reservoir with treatment is an important element of any program for \textit{P. vivax} elimination \cite{campo2015killing}. Mass drug administration (MDA) is an effective intervention for controlling many diseases and was advocated by the World Health Organization (WHO) in the 1950s to control malaria transmission \cite{hsiang2013mass}. MDA involves treating the entire population, or a well-defined sub-population, in a geographic location regardless of their infection status \cite{newby2015review,hsiang2013mass}. Most of the antimalarial drugs currently used to treat malaria only clear blood-stage parasites. Drugs that clear hypnozoites from the liver are referred to as `radical cure', examples of which are primaquine and tafenoquine  \cite{wells2010targeting,taylor2019short,poespoprodjo2022supervised}. In a radical cure MDA intervention, individuals are given a combination of two such drugs: artemisinin combination therapy (ACT) for clearing blood-stage parasites and primaquine to clear hypnozoites. However, because of the risk of haemolysis in glucose 6 phosphate dehydrogenase (G6PD)–deficient individuals, radical cure is not recommended by the WHO without screening for G6PD deficiency \cite{world2021second,howes2012g6pd, watson2018implications}. \par

The effect of radical cure treatment on \textit{P. vivax} transmission has been explored in a number of mathematical models \cite{ishikawa2003mathematical,aguas2012modeling,chamchod2013modeling,roy2013potential,white2014modelling,white2016variation,white2018mathematical}. However, most of these model do not consider the variation in hypnozoite number per mosquito bite \cite{ishikawa2003mathematical,aguas2012modeling,chamchod2013modeling,roy2013potential,white2016variation}. Mehra \textit{et al.} \cite{mehra2022hypnozoite} have developed a within-host model capturing hypnozoite dynamics and variation across infected mosquito bites that explicitly models the effect of radical cure treatment on the hypnozoite dynamics and reservoir. We have previously developed a multiscale model \cite{anwar2022multiscale} by embedding Mehra \textit{et al.}’s within-host model without treatment, which only uses three compartments at the population level while considering hypnozoite dynamics and the effect of the hypnozoite reservoir on disease transmission. The effect of three rounds of MDA with radical cure on \textit{P. vivax} prevalence has been studied in a randomised controlled trial \cite{phommasone2020mass}. However, as the MDA implementations are expensive, and the empirical evidence remains unclear as to the overall impact they are expected to have, mathematical modelling is well suited  to explore the overall impact and establish efficient designs before the actual implementation of the MDAs \cite{kaehler2019promise,jambulingam2016mathematical}. The impact of multiple rounds ($>3$) of MDA on \textit{P. vivax} transmission has not been explored with a mathematical model as far as we are aware. As \textit{P. vivax} is transmitted by mosquitoes, overall disease transmission is greatly affected by the mosquito population distribution in a region, which in turn is influenced by climate factors \cite{herdicho2021optimal,buonomo2018optimal,bashar2014seasonal,galardo2009seasonal}. Thus, the effect of treatment can be influenced by an abundance of mosquitoes and, hence, by seasonality. However, while a few \textit{P. vivax} transmission models have considered the role of seasonality in mosquito population dynamics \cite{ishikawa2003mathematical,chamchod2013modeling,silal2019malaria, mehra_2022_arxiv}, few have also captured the rich dynamics of hypnozoites \cite{mehra_2022_arxiv} and none have considered the impact of drug administration. Also, the abundance of mosquitoes and the contribution of hypnozoite activation can frequently trigger superinfection (reinfection of individuals that are already infected) which can potentially delay recovery from infection (\cite{smith2012ross,dietz1974malaria}). However, only a few \textit{P. vivax} mathematical transmission models account for superinfection \cite{white2014modelling,white2018mathematical,silal2019malaria, thesis_somya, mehra_2022_arxiv}.\par

In this article, we study the impact of multiple rounds of radical cure treatment within an MDA program on disease transmission by incorporating hypnozoite dynamics into an epidemic transmission framework. We account for superinfection and consider the impact of seasonal mosquito population changes (which we refer to as ``seasonality'' throughout). In Section \ref{model}, we extend our existing multiscale model \cite{anwar2022multiscale} to incorporate the effect of radical cure treatment and seasonality. We then obtain some key parameters for the population model from the within-host model \cite{mehra2022hypnozoite} under multiple rounds of MDA and obtain the recovery rate under superinfection. In Section \ref{result}, we provide illustrative results for both the within-host scale and transmission setting. We construct an optimisation problem to determine the optimal interval between MDA rounds with and without accounting for seasonality before concluding remarks are presented in Section \ref{disscussion}.

\section{Methods}\label{model}
A multiscale mathematical model that accounts for hypnozoite variation within individuals without treatment has already been developed \cite{anwar2022multiscale}. In our previous work, we did not account for superinfection in the population level model. Here, we extend the population level model to account for superinfection and allow treatment (via MDA) with a radical cure. The inclusion of superinfection in the model is important since for high transmission settings, overlapping blood-stage infections are frequent due to exposure to multiple infectious bites as well as the activation of hypnozoites.

\subsection{Population transmission model with treatment}
 
 Let $S$, $I$ and $L$ represent the fraction of the human population who are susceptible with no hypnozoites, blood-stage infected and liver-stage infected, respectively. Individuals in both $S$ and $L$ compartments are susceptible to infective mosquito bites and become blood-stage infected ($I$) at the rate $\lambda(t)=mabI_m$, where $m$ is the human-to-mosquito ratio, $a$ is mosquito biting rate, and $b$ is the transmission probability from mosquito to human. Recovery without accounting for superinfection is straightforward. In our previous model \cite{anwar2022multiscale}, we did not account for superinfection in the population-level model while the within-host framework permits superinfection. Here, we consider superinfection at the population level by following the work of Mehra \cite{thesis_somya}. To do that we need to consider the multiplicity of infection (MOI), defined as the number of distinct parasites co-circulating within a blood-stage infected individual (for \textit{P. vivax}, either from a new infectious bite or hypnozoite activation). When considering superinfection, an individual might experience multiple blood-stage infections, and recovery from the blood-stage infection is conditioned upon how many infections (MOI) they are currently experiencing. Those blood-stage infected individuals who are experiencing only one infection will recover and move out of $I$ and, depending on the hypnozoite reservoir size, either become susceptible ($S$) or liver-stage infected ($I$). Following the work of Mehra \cite{thesis_somya}, we define two parameters $p_1(t)$ and $p_2(t)$ where $p_1(t)$ is the probability that a blood-stage infected individual only experiencing one infection (MOI=1) has no hypnozoites in their liver at time $t$ and $p_2(t)$ is the probability that a blood-stage infected individual only experiencing one infection (MOI=1) has hypnozoites in their liver at time $t$. Hence, after recovery from blood-stage infection, individuals become susceptible ($S$) at rate $p_1(t)\gamma$ and become liver-stage infected ($L$) at rate $p_2(t)\gamma$, where $\gamma$ is the natural recovery rate. Thus, the probability of staying blood-stage infected at time $t$ is $\big(1-(p_1(t)+p_2(t))\big)$.

\begin{figure}[ht]
\centering
  \includegraphics[width=160mm]{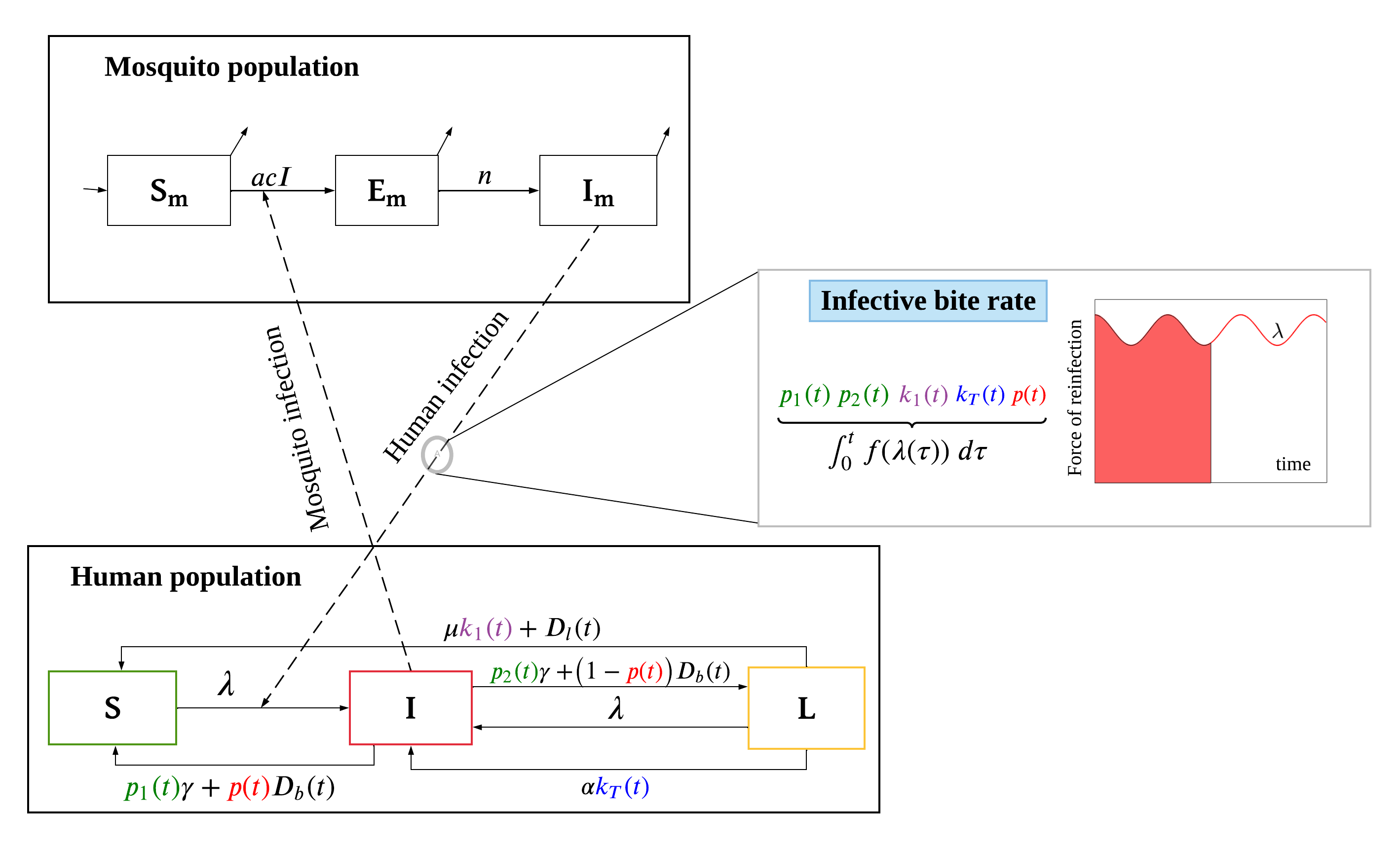}
  \caption{\textit{Schematic illustration of the multiscale model with treatment. $S$, $I$ and $L$ represent the fractions of the human population that are susceptible with no hypnozoites, blood-stage infected, and liver-stage infected, respectively. The left (top and bottom) part of the schematic demonstrates the transmission dynamics between the human and mosquito populations while the right part of the schematic demonstrates how the within-host model has been embedded within the population scale model. The within-host model takes into account the history of infective bites and calculates the probability of blood-stage infected individuals having $0$ hypnozoites and one blood-stage infection ($p_1(t)$), blood-stage infected individuals having more than $0$ hypnozoites and one blood-stage infection ($p_2(t)$), liver-stage infected individuals having 1 hypnozoite ($k_1(t)$), the expected size of the hypnozoite reservoir ($k_T(t)$), and the probability of blood-stage infected individuals having $0$ hypnozoites ($p(t)$) at any given time $t$ as a function of the force of infection, $\lambda(t)$. The red area on the right part of the schematic indicates the force of infection from time $t=0$ to $t$. The functions $D_b(t)$ and $D_l(t)$ capture the effect of treatment when implemented. Other parameters are defined in Table \ref{tab:white}.}}
  \label{fig:multiscale}
\end{figure}

 \begin{landscape}
\begin{table}[ht]
\centering
\footnotesize
\caption{Definitions, values and sources for model parameters\label{tab:white}.}
\begin{tabular}[t]{clcc}
\toprule
\textbf{Symbol} & \textbf{Definition} &  \textbf{Value/s} & \textbf{Source}  \\
\midrule
$a$ & Biting rate of mosquitoes & 80  year$^{-1}$ & \parencite{garrett1964human} \\
$b$ & Transmission probability: mosquito to human & 0.5& \parencite{smith2010quantitative} \\
$c$ & Transmission probability: human to mosquito & 0.23& \parencite{bharti2006experimental}\\
$b_m(t)$ & Mosquito birth rate (seasonal) & Time-varying &    \\
$g$ & Baseline mosquito birth rate & 0.1 day$^{-1}$& \parencite{gething2011modelling}   \\
$\eta$ & Seasonal amplitude & 0.1 & Assumed \\ 
$\phi$ & Seasonal phase & 0 & Assumed\\ 
$m$ & Number of mosquitoes per human & Varried&  \\ 
$n$ & Rate of mosquito sporogony & 1/12 days$^{-1}$ & \parencite{gething2011modelling} \\
$\gamma$ & Blood-stage infection clearance rate & 1/60 day$^{-1}$ [1/156 1/46]& \parencite{collins2003retrospective} \\
$\alpha$ & Hypnozoite activation rate & 1/332 day$^{-1}$ [1/700 1/110]& \parencite{white2014modelling}\\
$\mu$ & Hypnozoite death rate & 1/425 day$^{-1}$ [1/750 1/155]& \parencite{white2014modelling} \\
$\nu $ & Average number of hypnozoites per mosquito bite & 8.5 [1.5 20.5]& \parencite{white2014modelling}\\
$\lambda(t)$ & Force of reinfection & Calculated & $\lambda(t)=m_0abI_m$\\
$p_1(t)$ & Probability that a blood-stage infected individual has no hypnozoites and $MOI=1$& Time-varying &Calculated\\
$p_2(t)$ & Probability that a blood-stage infected individual has hypnozoites and $MOI=1$ & Time-varying &Calculated\\
$k_i(t)$ & Probability that liver-stage infected individual has $i$ hypnozoites within liver & Time varying&Calculated,  \parencite{mehra2022hypnozoite}\\
$k_T(t)$ & Average number of hypnozoites within liver for liver-stage infected individuals & Time varying&Calculated,  \parencite{mehra2022hypnozoite}\\
$p_{blood}$ & Probability that ongoing blood-stage infections are cleared instantaneously
 & 0.9 & Assumed\\
$p_{rad}$ & Probability that hypnozoites dies instantaneously & 0.9 & Assumed \\
$D_b(t)$ & Clearance rate of blood-stage parasite & Calculated & \\
$D_l(t)$ & Clearance rate of liver-stage parasite (hypnozoite) & Calculated & \\
$N$ & Total number of MDA rounds & varied & \\
$s_j (j=1,\ 2, \ldots, N)$ & MDA administration time & Estimated with optimisation& \\
$x_j (j=1,\ 2, \ldots, N-1)$ & MDA intervals & Estimated with optimisation& \\
\bottomrule
\end{tabular}
\end{table}
\end{landscape}

 Individuals suffer relapses from hypnozoite activation, the rate at which depends on the hypnozoite reservoir size and the baseline activation rate for each hypnozoite, $\alpha$. We define $k_T(t)$ to be the average hypnozoite reservoir size given liver-stage infected. That is $\alpha k_T(t)$ is the relapse rate. Individuals from the $L$ compartment become susceptible without experiencing a relapse if they have only one hypnozoite (with probability $k_1(t)$) and the hypnozoite dies naturally before activation, at rate $\mu$. For the mosquito population, we define $S_m,\ E_m$, and $I_m$ to be the fraction of susceptible, exposed, and infectious mosquitoes, respectively. Susceptible mosquitoes become exposed when they take a blood meal from an infected individual at the rate $acI$, where $c$ is the transmission probability from human to mosquito. After the incubation period (mean $1/n$ days), they become infectious and can transmit parasites to humans. The time-dependent parameters $p_1(t),\ p_2(t),\ k_1(t)$, and $k_T(t)$ capture the dynamics of hypnozoites and are obtained from the within-host model (Section \ref{WH}) as an integral function of the force of infection, $\lambda(t)$, which makes the model a system of integro-differential equations (IDEs). The definition and derivation of these time-dependent parameters are discussed in Section \ref{WH}. The model schematic is depicted in Figure \ref{fig:multiscale}. \par

 Suppose that drug treatment is administered successively at times $s_1, s_2,\ldots, s_N$, where $N$ is the total number of MDA rounds. The effect of blood-stage and liver-stage radical cure treatments are captured by the time-dependent functions $D_b(t)$ and $D_l(t)$, respectively. To implement the effect of radical cure in the population level model, we assume that radical cure has an instantaneous effect \cite{mehra2022hypnozoite}. That is, on administration, all ongoing blood-stage infections are instantaneously cleared with probability $p_{blood}$, and each hypnozoite in the liver dies instantaneously with probability $p_{rad}$. Without treatment, blood-stage infections are cleared one at a time, but with treatment, blood-stage infections will all be cleared with probability $p_{blood}$. We define $p(t)$ as the probability of blood-stage infected individuals having no hypnozoites in their liver \cite{anwar2022multiscale}. Therefore, at the time when radical cure is administered, individuals that were blood-stage infected either become susceptible with probability $p(t)$ or become liver-stage infected with probability $(1-p(t))$. Blood-stage infected individuals who are not cured following treatment undergo the same dynamics as those who receive no treatment. Liver-stage infected individuals whose hypnozoites have not been fully cleared following treatment will undergo the same dynamics as if without treatment but starting with the reduced hypnozoite reservoir. Hence, if radical cure is administered at time $t=s_j$, where $j$ is the number of MDA rounds, the drug has an effect only on the ongoing infections and hypnozoites established from time $t=s_{j-1}$ until $t=s_j$. Hypnozoites that are established after time $t=s_j$ or any blood-stage infections caused by either hypnozoite activation or infectious mosquito bites after $t=s_1$ will undergo dynamics as if without treatment (until the next time of MDA application). Since we are concerned with disease dynamics over a time scale of years, the assumption of an instantaneous effect of the radical cure is appropriate, as drugs such as artemisinin, which clears blood-stage parasites, have a half-life of 1.93 hours \cite{birgersson2016population} and primaquine, which kills hypnozoites have a relatively short half-life of approximately 3.7\mbox{--}9.6 hours \cite{jittamala2015pharmacokinetic}. Another drug, tafenoquine, that also kills hypnozoites has a half-life of approximately 14\mbox{--}28 days which is short compared to a time scale of years \cite{SCHLAGENHAUF2019145}. Since the number of mosquitoes in the environment influences \textit{P. vivax} dynamics dramatically \cite{herdicho2021optimal,buonomo2018optimal}, it is important to account for seasonal environmental effects on the mosquito population (see, for example, \cite{bashar2014seasonal,galardo2009seasonal}). To incorporate  mosquito seasonality, we consider that the mosquito birth rate at time t, $b_m(t)$, is regulated by a cosine function with a period of 1 year as follows:

\begin{align*}
    b_m(t)=b_m(0)\left(1+\eta \cos\left(\frac{2\pi t}{365}+\phi\right)\right),
\end{align*}
where $b_m(0)=g$ is the baseline mosquito birth rate, $\eta\in[0\ 1)$ is the seasonal amplitude and $\phi$ is the seasonal phase (taken to be 0). Note that if $b_m(t)=b_m(0)=g$, then the mosquito population is constant, that is, without seasonality. With all the assumptions outlined above, the system of IDEs that describe the dynamics is (see Appendix \ref{prop_model} for a detailed derivation of the model):
\begin{align}
    &\frac{\mathrm{d}S}{\mathrm{d}t}=-\lambda S+\mu k_1(t)L+p_1(t)\gamma I+D_l(l)L+D_b(t)p(t)I,\\
    \label{eqn:pp2}
     &\frac{\mathrm{d}I}{\mathrm{d}t}=\lambda(S+I)+\alpha k_T(t)L-\gamma \big(p_1(t)+p_2(t)\big) I-D_b(t)I,\\
    \label{eqn:pp3}
    &\frac{\mathrm{d}L}{\mathrm{d}t}=-\lambda L-\mu k_1(t)L-\alpha k_T(t)L+\gamma p_2(t) I-D_l(t)L+D_b(t)\big(1-p(t)\big)I,\\
     &\frac{\mathrm{d}S_m}{\mathrm{d}t}=b_m(t)-acIS_m-b_m(t)S_m,\\
       \label{eqn:pp5}
   &\frac{\mathrm{d}E_m}{\mathrm{d}t}=acIS_m-\left(b_m(t)+n\right)E_m, \\
    \label{eqn:pp6}
    &\frac{\mathrm{d}I_m}{\mathrm{d}t}=nE_m-b_m(t)I_m,
\end{align}
where, 
\begin{align}
    \lambda=&m_0ab I_m  \text{exp}\left\{\frac{365g\eta}{2\pi}  \sin\left(\frac{2\pi t}{365}+\phi\right)\right\}, \nonumber
\end{align}
is the force of reinfection, and $m_0=\frac{N_m(0)}{N_h}$ is the initial mosquito ratio. Here $D_b(t)$ and $D_l(t)$ are blood-stage parasite and liver-stage parasite (hypnozoite) clearance rates respectively and are given by:

\begin{align*}
D_b(t)=&\ln{\big((1-p_{blood})^{-1}\big)}\sum_{j=1}^N \delta_D(t-s_j),\\
  D_l(t)=&\big\{k_1(t)\ln{\big((1-p_{rad})^{-1}}\big)+k_2(t)\ln{\big((1-p_{rad}^2)^{-1}}\big)+\ldots\big\}\sum_{j=1}^N \delta_D(t-s_j)\\
   =&\sum_{i=1}^\infty \ln{\Big({1-p_{rad}^i}}\Big)^{-k_i(t)} \sum_{j=1}^N \delta_D(t-s_j),
\end{align*}
where $\delta_D(\cdot)$ is the Dirac delta function. That is, any blood-stage parasite will be instantaneously cleared with probability $p_{blood}$ every time the radical cure is administered \cite{mehra2022hypnozoite} and depending on the parameter $p_1(t)$ which is the probability that an individual experiencing only one infection has no hypnozoites in the liver given blood-stage infection (Equation \ref{eqn:p_1_final}) and $p_2(t)$ which is the probability that an individual experiencing only one infection has hypnozoites in the liver given blood-stage infection (Equation \ref{eqn:p_2}), move to the susceptible compartment ($S$) at rate $p_1(t)D_b(t)$ and to the liver-stage infected compartment ($L$) at rate $p_2(t)D_b(t)$, respectively. As each hypnozoite is cleared with probability $p_{rad}$, the liver-stage clearance rate $D_l(t)$ depends on how many hypnozoites are present in the liver.  That is, $D_l(t)$ depends on $k_1(t),\ k_2(t),\ \ldots, k_i(t)$, where $k_i(t)$ is the probability that a liver-stage infected individual has $i$ hypnozoites. All model parameters are defined in Table \ref{tab:white}.

\subsection{Within-host model with treatment} \label{WH}
 A within-host model considering the effect of radical cure on hypnozoite dynamics was introduced by Mehra \textit{et al.} \cite{mehra2022hypnozoite}. They developed the framework considering $N$ MDA rounds but explored analytically and numerically considering one MDA round. Here, we solve the necessary equations for $N$ MDA rounds. First,  the dynamics of a single hypnozoite under treatment were modelled, then a fixed number of hypnozoites introduced by a single mosquito bite before accounting for continuous mosquito inoculation where each mosquito bite contributes an average of $\nu$ hypnozoites to the reservoir. The within-host model also assumes that radical cure has an instantaneous effect. \par

For the short-latency case (in which hypnozoites can immediately activate after establishment without going through a latency phase), a hypnozoite can be in one of four different states. Let $H$, $A$, $C$, and $D$ represent the state of establishment, activation, clearance and death for a single hypnozoite, respectively. Suppose that drug treatment is administered successively at times $s_1, s_2,\ldots, s_N$. We denote the state of the hypnozoite at time $t$ with
$X_r(t,\ s_1,\ s_2,\ \ldots, s_N)\in (H,A,C,D)$ with corresponding probability mass function (PMF) $$p^r_H(t,s_1,\ldots,s_N),\ p^r_A(t,s_1,\ldots,s_N),\ p^r_C(t,s_1,\ldots,s_N),\ p^r_D(t,s_1,\ldots,s_N)$$ respectively. The governing equations for the state probabilities under treatment are given by Equations (17)--(22) from Mehra \textit{et al.}\ \cite{mehra2022hypnozoite}:

\begin{align}
\label{eqn:1}
    \frac {dp^r_H}{dt}=&-(\alpha+\mu)p^r_H-\ln{\big((1-p_{\text{rad}})\big)^{-1}}\sum_{j=1}^N \delta_D(t-s_j)p^r_H,\\
    \label{eqn:2}
    \frac {dp^r_A}{dt}=&-\gamma p^r_A+\alpha p^r_H-\ln{\big((1-p_{\text{blood}})\big)^{-1}}\sum_{j=1}^N \delta_D(t-s_j)p^r_A,\\
    \label{eqn:3}
    \frac {dp^r_C}{dt}=&\gamma p^r_A+\ln{\big((1-p_{\text{blood}})\big)^{-1}}\sum_{j=1}^N \delta_D(t-s_j)p^r_A,\\
    \label{eqn:4}
    \frac {dp^r_D}{dt}=&-\mu p^r_H+\ln{\big((1-p_{\text{rad}})\big)^{-1}}\sum_{j=1}^N \delta_D(t-s_j)p^r_H,
\end{align}
where the parameters $\alpha,\ \gamma,$ and $\mu$ are as per Table \ref{tab:white}.
  Since our population model in Equations (\ref{eqn:pp2})--(\ref{eqn:pp6}) uses the parameters $p_1(t),\ p_2(t),\ k_1(t)$ and $k_T(t)$, we seek to obtain expressions for these parameters from the within-host model under multiple rounds of MDA. Evaluating the parameters $p_1(t),\ p_2(t),\ k_1(t)$ and $k_T(t)$ in the population model requires the probability of hypnozoite establishment ($p^r_H(t)$) and the probability of hypnozoite activation ($p^r_A(t)$) \cite{anwar2022multiscale}; hence we solve Equations (\ref{eqn:1})--(\ref{eqn:2}) for $N$ MDA rounds to give:
 
 \begin{align}
 \label{eqn:pH_r}
    p^r_H(t,s_1,s_2,s_3,\ldots, s_N)=&(1-p_{\text{rad}})^N p_H(t),\\
    \label{eqn:pA_r}
    p^r_A(t,s_1,s_2,s_3,\ldots, s_N)=&(1-p_{\text{blood}})e^{-\gamma(t-s_N)}p_A^r(s_N,s_1,s_2,\ldots,s_{N-1})\nonumber\\  &+(1-p_{\text{rad}})^N\big(p_A(t)-e^{-\gamma(t-s_N)}p_A(s_N)\big),
\end{align}
where $p_H(t)$ and $p_A(t)$ are the probability of establishment and activation of a hypnozoite without treatment, respectively, and are given by: 

\begin{align*}
    p_H(t)=&e^{-(\alpha +\mu)t},\\
    p_A(t)=&\frac{\alpha}{(\alpha+\mu)-\gamma}\left(e^{-\gamma t}-e^{-(\alpha +\mu)t}\right).
\end{align*}

\begin{figure}[htp]
    \begin{centering}
        \includegraphics[width=1\textwidth]{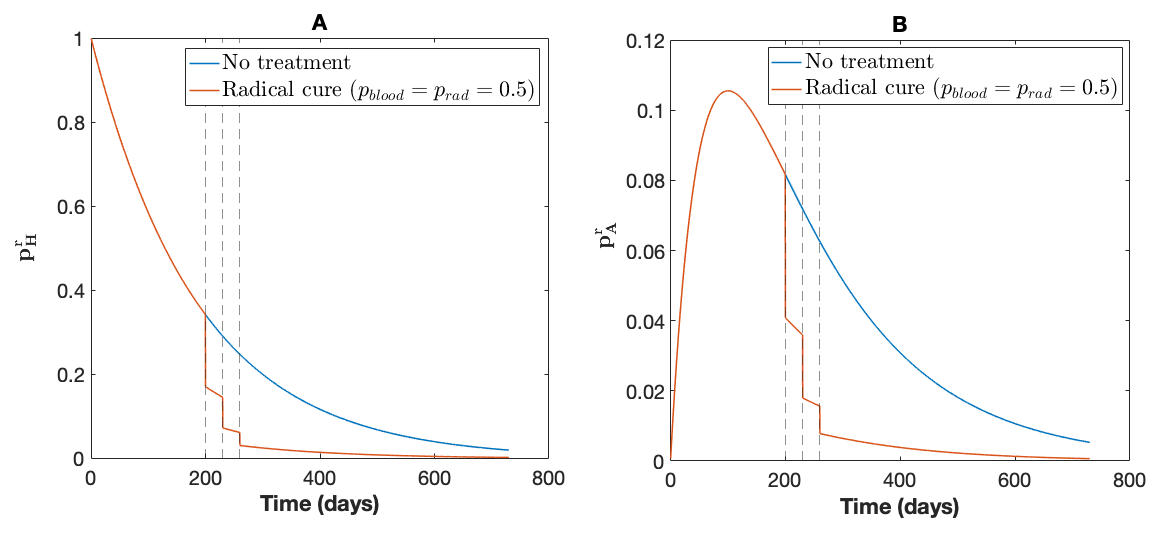}
    \caption{\textit{Effect of radical cure (three rounds of MDA) on a single hypnozoite. (A) Probability of hypnozoite establishment as per Equation (\ref{eqn:pH_r}) and (B) probability of hypnozoite activation as per Equation (\ref{eqn:pA_r}). For each subplot, blue represents the probability without considering treatment and orange represents treatment assuming 50\% efficacy $(p_{blood}=p_{rad}=0.5)$ of the drugs. The vertical lines indicate the times of drug administration. Other parameters are as in Table \ref{tab:white}.}}
    \label{fig:pA_pH}
    \end{centering}
\end{figure}

Figure \ref{fig:pA_pH} shows the effect of three rounds of MDA on the dynamics of a single hypnozoite. The probability of hypnozoite establishment ($p^r_H(t)$) and hypnozoite activation ($p^r_A(t)$) under 3 rounds of MDA (with $p_{blood}=p_{rad}=0.5$) are illustrated in Figure \ref{fig:pA_pH}(A) and Figure \ref{fig:pA_pH}(B), respectively. The drug is administered for the first time 200 days after the hypnozoite is established and the interval between each MDA round is fixed at 30 days.

We now define two additional states, $P$ and $PC$, to denote an ongoing primary infection from infective mosquito bites and a cleared primary infection, respectively. 
Let $N_f(t)$ denote the number of hypnozoites in states  $f\in \{H,A,C,D\}:=F$ at time $t$ and $N_P(t),\ N_{PC}(t)$ denote the number of ongoing and cleared primary infections, respectively, at time $t$. Defining the state space $F':=\left\{H,A,C,D,P,PC\right\}$, the probability generating function (PGF) for 

$$\mathbf{N} (t)=(N_H(t),N_A(t),N_C(t),N_D(t),N_P(t),N_{PC})$$

with $\mathbf{N} (0)=\mathbf{0}$ can be written following from Equation (30) in Mehra \textit{et al.} \parencite{mehra2022hypnozoite} (for short-latency case ($k=0$) with probability of a blood-stage infection after an infectious bite, $p_{prim}=1$) (by the law of total expectation):
\begin{align}
    \label{PGF_ntrt}   G(t,z_H,z_A,z_C,z_D,z_P,z_{PC}):&=\mathbb{E}\left[\displaystyle\prod_{f\in F'} z_f^{N_f(t)}\right]\\ \nonumber&=\text{exp}\left\{-q(t)+\int_0^t \frac{\lambda(\tau)\left(z_Pe^{-\gamma (t-\tau)}+(1-e^{-\gamma (t-\tau)})z_{PC}\right)}{1+\nu \left(1-\sum_{f\in F} z_fp_f(t-\tau)\right)}d\tau \right\},
    \end{align}
where $q(t)$ is the mean number of infective bites in the interval $(0,t]$ and is given by: 
    \begin{align*}
        q(t)=\int_0^t \lambda(\tau)d\tau.
    \end{align*}
All parameters are as per Table \ref{tab:white}. The expression for the joint PGF with drug  administration at time $t=s_1$ is given by Equation (31) in Mehra \textit{et al.} \cite{mehra2022hypnozoite}. Following a similar analysis, if the drug is administered at $N$ successive times ($s_1,\ s_2,\ldots, s_N$) then the joint PGF for the number of hypnozoites/infections in each state is: \begin{align}
    &G^{s_1,s_2,\ldots s_N}(t,z_H,z_A,z_C,z_D,z_P,z_{PC}):=\mathbb{E}\left[\displaystyle\prod_{f\in F'} z_f^{N^{s_1,s_2,\ldots s_N}_s(t)}\right] \nonumber\\ 
    &=\text{exp}\Bigg\{-q(t)+\int_{s_N}^t \lambda(\tau)\frac{e^{-\gamma (t-\tau)}z_P+(1-e^{-\gamma (t-\tau)})z_{PC}}{1+\nu \left(1-\sum_{f\in F} z_f.p_f(t-\tau)\right)}d\tau  \nonumber\\ 
    &  +\int_{0}^{s_1} \lambda(\tau)\frac{(1-p_{\text{blood}})e^{-\gamma (t-\tau)}z_P+(1-(1-p_{\text{blood}})e^{-\gamma (t-\tau)})z_{PC}}{1+\nu \left(1-\sum_{f\in F} z_f.p^r_s(t-\tau,s_1-\tau)\right)}d\tau \nonumber \\& +\int_{s_1}^{s_2} \lambda(\tau)\frac{(1-p_{\text{blood}})^2e^{-\gamma (t-\tau)}z_P+(1-(1-p_{\text{blood}})^2e^{-\gamma (t-\tau)})z_{PC}}{1+\nu \left(1-\sum_{f\in F} z_f.p^r_s(t-\tau,s_1-\tau,s_2-\tau)\right)}d\tau \nonumber\\ & +\int_{s_2}^{s_3} \lambda(\tau)\frac{(1-p_{\text{blood}})^3e^{-\gamma (t-\tau)}z_P+(1-(1-p_{\text{blood}})^3e^{-\gamma (t-\tau)})z_{PC}}{1+\nu \left(1-\sum_{f\in F} z_f.p^r_s(t-\tau,s_1-\tau,s_2-\tau,s_3-\tau)\right)}d\tau \nonumber\\ & +\ldots +\int_{s_{N-1}}^{s_N} \lambda(\tau)\frac{(1-p_{\text{blood}})^Ne^{-\gamma (t-\tau)}z_P+(1-(1-p_{\text{blood}})^Ne^{-\gamma (t-\tau)})z_{PC}}{1+\nu \left(1-\sum_{f\in F} z_f.p^r_s(t-\tau,s_1-\tau,s_2-\tau,s_3-\tau,\ldots, s_N-\tau)\right)}d\tau \Bigg\}.
    \label{PGF}
\end{align}

We now use the PGF in Equation (\ref{PGF}) to derive expressions for the population-level parameters $p(t)$, $p_1(t),\ p_2(t),\ k_1(t)$, and $k_T(t)$ under multiple MDA rounds.

\subsubsection{Probability blood-stage infected individual has no hypnozoites (under \texorpdfstring{$N$}{} rounds of MDA)}
 With $p(t)$ defined as the probability that an individual has an empty hypnozoite reservoir conditional on an ongoing blood-stage infection (i.e. primary infection or relapse) from Equation (13) of Anwar \textit{et al.} \cite{anwar2022multiscale} we have:

\begin{align}
\label{eqn:3.12}
  p(t)&=P\big(N_H(t)=0|N_A(t)>0 \cup N_P(t)>0\big)\nonumber\\
  &=\frac{P\big(N_H(t)=0)-P(N_H(t)=N_A(t)=N_P(t)=0\big)}{1-P\big(N_A(t)=N_P(t)=0\big)}.
\end{align}

where the probability that an individual has an empty hypnozoite reservoir at time $t$, $P(N_H(t)=0)$, is given by:
\begin{align}
  P(&N_H(t)=0)=G^{t,s_1,s_2,\ldots,s_N}(t,z_H=0, z_A=1, z_C=1, z_D=1, z_P=1, z_{PC}=1)\nonumber\\
  \label{eqn:3.13}
  &=
  \begin{cases}
    \text{exp}\Bigl\{-q(t)+\int_0^t \frac{\lambda(\tau)}{1+\nu p_H(t-\tau)}d\tau \Bigr\}& \text{if}\ t< s_1\\ 
    \text{exp}\Bigl\{-q(t)+\int_{s_N}^t \frac{\lambda(\tau)}{1+\nu p_H(t-\tau)}d\tau+\int_0^{s_1} \frac{\lambda(\tau)}{1+\nu p_H^r(t-\tau,s_1-\tau)}d\tau\\\qquad +\int_{s_1}^{s_2} \frac{\lambda(\tau)}{1+\nu p_H^r(t-\tau,s_1-\tau,s_2-\tau)}d\tau+\ldots +\int_{s_{N-1}}^{s_N} \frac{\lambda(\tau)}{1+\nu p_H^r(t-\tau,s_1-\tau,\ldots,s_N-\tau)}d\tau \Bigr\}& \text{if}\ t\ge s_N,
    \end{cases} 
\end{align}
the probability that an individual is neither experiencing a relapse nor a primary infection at time $t$, $P\big(N_A(t)+N_P(t)=0\big)$ (i.e., no blood-stage infection), is given by:
\begin{align}
  P\big(N_A(t)+N_P(t)=0\big)=&G^{t,s_1,s_2,\ldots,s_N}(t,z_H=1, z_A=0, z_C=1, z_D=1, z_P=0, z_{PC}=1) \nonumber \\
  =&
  \begin{cases}
  \text{exp}\Bigl\{-q(t)+\int_0^t \frac{\lambda(\tau)(1-e^{-\gamma (t-\tau)})}{1+\nu p_A(t-\tau)}d\tau\Bigr\} & \text{if}\ t< s_1\\
  \text{exp}\Bigl\{-q(t)+\int_{s_N}^t \frac{\lambda(\tau)(1-e^{-\gamma (t-\tau)})}{1+\nu p_A(t-\tau)}d\tau\\\qquad+\int_0^{s_1} \frac{\lambda(\tau)(1-(1-p_{blood})e^{-\gamma (t-\tau)})}{1+\nu p_A^r(t-\tau,s_1-\tau)}d\tau\\\qquad+\int_{s_1}^{s_2} \frac{\lambda(\tau)(1-(1-p_{blood})^2e^{-\gamma (t-\tau)})}{1+\nu p_A^r(t-\tau,s_1-\tau,s_2-\tau)}d\tau\\\qquad+\ldots+\int_{s_{N-1}}^{s_N} \frac{\lambda(\tau)(1-(1-p_{blood})^Ne^{-\gamma (t-\tau)})}{1+\nu p_A^r(t-\tau,s_1-\tau,\ldots,s_N-\tau)}d\tau\Bigr\} & \text{if}\ t\ge s_N.
  \end{cases}
  \label{eqn:3.14}
\end{align}

and
the probability that an individual is neither experiencing an
infection nor has any hypnozoites in their liver at time $t$, $P\big(N_H(t)=N_A(t)=N_P(t)=0\big)$, is given by:

\begin{align}
P\big(N_H(t)=&N_A(t)=N_P(t)=0\big)\nonumber\\=&G^{t,s_1,s_2,\ldots,s_N}(t,z_H=0, z_A=0, z_C=1, z_D=1, z_P=0, z_{PC}=1)\nonumber\\
  =&\begin{cases}
  \text{exp}\Bigl\{-q(t)+\int_0^t \frac{\lambda(\tau)(1-e^{-\gamma (t-\tau)})}{1+\nu(p_H(t-\tau)+p_A(t-\tau)}d\tau\Bigr\} & \text{if}\ t< s_1\\
  \text{exp}\Bigl\{-q(t)+\int_{s_N}^t \frac{\lambda(\tau)(1-e^{-\gamma (t-\tau)})}{1+\nu(p_H(t-\tau)+p_A(t-\tau)}d\tau
  \\\qquad +\int_0^{s_1} \frac{\lambda(\tau)(1-(1-p_{blood})e^{-\gamma (t-\tau)})}{1+\nu(p_H^r(t-\tau,s_1-\tau)+p_A^r(t-\tau,s_1-\tau)}d\tau\\\qquad +\int_{s_1}^{s_2} \frac{\lambda(\tau)(1-(1-p_{blood})^2e^{-\gamma (t-\tau)})}{1+\nu(p_H^r(t-\tau ,s_1-\tau,s_2-\tau)+p_A^r(t-\tau ,s_1-\tau,s_2-\tau)}d\tau \\\qquad +\ldots+\int_{s_{N-1}}^{s_N} \frac{\lambda(\tau)(1-(1-p_{blood})^Ne^{-\gamma (t-\tau)})}{1+\nu(p_H^r(t-\tau ,s_1-\tau,\ldots ,s_N-\tau)+p_A^r(t-\tau ,s_1-\tau,\ldots ,s_N-\tau)}d\tau\Bigr\}  & \text{if}\ t\ge s_N.
  \end{cases}
  \label{eqn:3.15}
\end{align}

\subsubsection{Probability of blood-stage infected individual having one infection and no hypnozoites (under \texorpdfstring{$N$}{} rounds of MDA)}
 With $p_1(t)$ defined as the probability that an individual has one infection ($N_A(t)+N_P(t)=1$) and an empty hypnozoite reservoir ($N_H(t)=0$) conditional on an ongoing blood-stage infection (i.e. primary infections or relapse, $N_A(t)+N_P(t)>1$), we have:

\begin{align}
\label{eqn:p_1}
    p_1(t)
             =&\frac{P\big(N_A(t)+N_P(t)=1|N_H(t)=0\big)P(N_H(t)=0)}{1-P(N_A(t)+N_P(t)=0)}.
\end{align}

The expression for $P(N_H(t)=0)$ and $P(N_H(t)+N_P(t)=0)$
follows from Equations (\ref{eqn:3.13}) and (\ref{eqn:3.14}). The expression for $P\big(N_A(t)+N_P(t)=1|N_H(t)=0\big)$ can be obtained from Equation (\ref{eqn:h}) (see appendix \ref{MOI_I} for details) which is

\begin{align*}
P(N_A(t)+N_P(t)=&1|N_H(t)=0)=\text{exp}\left\{h(0,t)-h(1,t)\right\}\frac{\partial h(0,t)}{\partial z}\\
=&\frac{G(t,z_H=0,\ z_A=0,\ z_C=1,\ z_D=1,\ z_P=0\, z_{PC}=1)}{G(t,z_H=0,\ z_A=1,\ z_C=1,\ z_D=1,\ z_P=1\, z_{PC}=1)}\frac{\partial h(0,t)}{\partial z},\\
=&\frac{P(N_H(t)=N_A(t)=N_P(t)=0)}{P(N_H(t)=0)}\frac{\partial h(0,t)}{\partial z}.
 \end{align*}

Finally, from Equation (\ref{eqn:p_1}),
\begin{align}
\label{eqn:p_1_final}
        p_1(t)=&\frac{P\big(N_A(t)+N_P(t)=1|N_H(t)=0\big)P(N_H(t)=0)}{1-P(N_A(t)+N_P(t)=0)},\nonumber\\
        =&\frac{P(N_H(t)=N_A(t)=N_P(t)=0)}{1-P(N_A(t)=N_P(t)=0)}\frac{\partial h(0,t)}{\partial z},
\end{align}
where 

\begin{align*}
\frac{\partial h(0,t)}{\partial z}=&\begin{cases}
\int_0^t \lambda(\tau)\frac{e^{-\gamma (t-\tau)}\big(1+\nu p_H(t-\tau)\big)+\nu p_A(t-\tau)}{\big[1+\nu\big(p_A(t-\tau)+p_H(t-\tau)\big)\big]^2}d\tau & \text{if}\ t< s_1\\
    \int_{s_N}^t \lambda(\tau)\frac{e^{-\gamma (t-\tau)}\big(1+\nu p_H(t-\tau)\big)+\nu p_A(t-\tau)}{\big[1+\nu\big(p_A(t-\tau)+p_H(t-\tau)\big)\big]^2}d\tau \\+\int_0^{s_1} \lambda(\tau)\frac{(1-p_{blood})e^{-\gamma (t-\tau)}\big(1+\nu p_H^r(t-\tau,s_1-\tau)\big)+\nu p_A^r(t-\tau,s_1-\tau)}{\big[1+\nu\big(p_A^r(t-\tau,s_1-\tau)+p_H^r(t-\tau,s_1-\tau)\big)\big]^2}d\tau \\ \int_{s_{1}}^{s_2} \lambda(\tau)\frac{(1-p_{blood})^2e^{-\gamma (t-\tau)}\big(1+\nu p_H^r(t-\tau,s_1-\tau,s_2-\tau)\big)+\nu p_A^r(t-\tau,s_1-\tau,s_2-\tau)}{\big[1+\nu\big(p_A^r(t-\tau,s_1-\tau,s_N-\tau)+p_H^r(t-\tau,s_1-\tau,s_N-\tau)\big)\big]^2}d\tau\\+\ldots
    +\int_{s_{N-1}}^{s_N} \lambda(\tau)\frac{(1-p_{blood})^Ne^{-\gamma (t-\tau)}\big(1+\nu p_H^r(t-\tau,s_1-\tau,\ldots,s_n-\tau)\big)+\nu p_A^r(t-\tau,s_1-\tau,\ldots,s_N-\tau)}{\big[1+\nu\big(p_A^r(t-\tau,s_1-\tau,\ldots,s_N-\tau)+p_H^r(t-\tau,s_1-\tau,\ldots,s_N-\tau)\big)\big]^2}d\tau & \text{if}\ t\ge s_N.
    \end{cases}
\end{align*}

\subsubsection{Probability of blood-stage infected individual having one infection and  non-zero hypnozoites (under \texorpdfstring{$N$}{} rounds of MDA)}

The probability that a blood-stage infected individual experiencing only one infection ($N_A(t)+N_P(t)=1$) and has hypnozoites ($N_H(t)>0$) at time $t$, $p_2(t)$, is

\begin{align}
\label{eqn:p_2}
    p_2(t)=&P(N_A(t)+N_P(t)=1,N_H(t)>0|P(N_A(t)+N_P(t)>0),\nonumber\\
             =&\frac{P\big(N_A(t)+N_P(t)=1\big)}{1-P(N_A(t)+N_P(t)=0)}-\frac{P\big(N_A(t)+N_P(t)=1|N_H(t)=0\big)P(N_H(t)=0)}{1-P(N_A(t)+N_P(t)=0)},\nonumber\\
    =&\frac{P\big(N_A(t)+N_P(t)=1\big)}{1-P(N_A(t)=N_P(t)=0)}-p_1(t).
\end{align}
The expression $P(N_A(t)+N_P(t)=0)=P(N_A(t)=N_P(t)=0)$ is given by Equation (\ref{eqn:3.14}). The expression for $P(N_A(t)+N_P(t)=1)$ follows from Equation (81) in Mehra \textit{et al.} \cite{mehra2022hypnozoite} and is given by

\begin{align*}
P(N_A(t)+N_P(t)=1)=&P(N_A(t)=N_P(t)=0)\frac{\partial f(0,t)}{\partial z},\\
 \end{align*}
where,
\begin{align*}
\frac{\partial f(0,t)}{\partial z}=&\begin{cases}
\int_0^t \lambda(\tau)\frac{e^{-\gamma (t-\tau)}+\nu p_A(t-\tau)}{[1+\nu p_A(t-\tau)]^2}d\tau & \text{if}\ t< s_1\\
    \int_{s_n}^t \lambda(\tau)\frac{e^{-\gamma (t-\tau)}+\nu p_A(t-\tau)}{[1+\nu p_A(t-\tau)]^2}d\tau \\+\int_0^{s_1} \lambda(\tau)\frac{(1-p_{blood})e^{-\gamma (t-\tau)}+\nu p_A^r(t-\tau,s_1-\tau)}{[1+\nu p_A^r(t-\tau,s_1-\tau)]^2}d\tau\\+\int_{s_{1}}^{s_2} \lambda(\tau)\frac{(1-p_{blood})^2e^{-\gamma (t-\tau)}+\nu p_A^r(t-\tau,s_1-\tau,s_2-\tau)}{[1+\nu p_A^r(t-\tau,s_1-\tau,s_2-\tau)]^2}d\tau \\ +\ldots
    +\int_{s_{N-1}}^{s_N} \lambda(\tau)\frac{(1-p_{blood})^Ne^{-\gamma (t-\tau)}+\nu p_A^r(t-\tau,s_1-\tau,\ldots,s_n-\tau)}{[1+\nu p_A^r(t-\tau,s_1-\tau,\ldots,s_n-\tau)]^2}d\tau & \text{if}\ t\ge s_N.
    \end{cases}
\end{align*}

\subsubsection{Probability liver-stage infected individual has 1 hypnozoite in liver (under \texorpdfstring{$N$}{} rounds of MDA)}\label{sec:i_hyp}

The probability that a liver-stage infected individual has 1 hypnozoite in the liver at time $t$ (that is, the conditional probability for $N_H(t)$ given an individual
does not have an ongoing blood-stage infection at time $t$) under $N$ MDA rounds is:

\begin{align}
k_1(t)=&P(N_H(t)=1|N_A(t)=N_P(t)=0,N_H(t)>0)\nonumber\\
 =&\frac{P(N_H(t)=1|N_A(t)=N_p(t)=0)}{1-P(N_H(t)=0|N_A(t)=N_P(t)=0)}.\nonumber\\
 \label{eqn:k1}
 =&\frac{\text{exp}\left\{g(0,t)-g(1,t)\right\}}{1-P(N_H(t)=0|N_A(t)=N_P(t)=0)} \frac{\partial g(0,t)}{\partial z}\nonumber\\
 =&\frac{P(N_H(t)=N_A(t)=N_P(t)=0)}{\big(1-P(N_H(t)=0|N_A(t)=N_P(t)=0)\big)P(N_A(t)=N_P(t)=0)} \frac{\partial g(0,t)}{\partial z},
 \end{align}

where 
\begin{align}
    \frac{\partial h(0,t)}{\partial z}=&\begin{cases}
    \int_0^t \frac{\lambda(\tau)\nu p_H(t-\tau)(1-e^{-\gamma (t-\tau)})}{[1+\nu(p_H(t-\tau)+p_A(t-\tau))]^2}d\tau & \text{if}\ t< s_1\\
    \int_{s_N}^t \frac{\lambda(\tau)\nu p_H(t-\tau)(1-e^{-\gamma (t-\tau)})}{[1+\nu(p_H(t-\tau)+p_A(t-\tau))]^2}d\tau \\ +\int_0^{s_1} \frac{\lambda(\tau)\nu p_H^r(t-\tau,s_1-\tau)(1-(1-p_{blood})e^{-\gamma (t-\tau)})}{[1+\nu(p_H^r(t-\tau,s_1-\tau)+p_A^r(t-\tau,s_1-\tau))]^2}d\tau\\ +\int_{s_{1}}^{s_2} \frac{\lambda(\tau)\nu p_H^r(t-\tau,s_1-\tau,s_2-\tau)(1-(1-p_{blood})^2e^{-\gamma (t-\tau)})}{[1+\nu(p_H^r(t-\tau,s_1-\tau,s_2-\tau)+p_A^r(t-\tau,s_1-\tau,s_2-\tau))]^2}d\tau\\ +\ldots
    +\int_{s_{N-1}}^{s_N} \frac{\lambda(\tau)\nu p_H^r(t-\tau,s_1-\tau,\ldots,s_N-\tau)(1-(1-p_{blood})^Ne^{-\gamma (t-\tau)})}{[1+\nu(p_H^r(t-\tau,s_1-\tau,\ldots,s_N-\tau)+p_A^r(t-\tau,s_1-\tau,\ldots,s_N-\tau))]^2}d\tau & \text{if}\ t\ge s_N.
    \end{cases}
\end{align}
The expression for $P(N_H(t)=1|N_A(t)=N_p(t)=0)$ 
follows from Equation (78) in \parencite{mehra2022hypnozoite} and $P(N_H(t)=0|N_A(t)=N_P(t)=0)$ is obtained by dividing Equation (\ref{eqn:3.15}) by Equation (\ref{eqn:3.14}).

\subsubsection{Average number hypnozoites within liver-stage infected individuals (under \texorpdfstring{$N$}{} rounds of MDA)}\label{sec:k_T}

The average number of hypnozoites within liver-stage infected individuals, $k_T(t)$, is defined by:
\begin{align}
\label{eqn:inf_sum}
   k_T=\sum_{i=1}^\infty ik_i &= \Big(\frac{\mathbb{E}\left[N_H(t)|N_A(t)=N_P(t)=0\right]}{1-P(N_H(t)=0|N_A(t)=N_P(t)=0)}\Big)\nonumber
\end{align}
where $\mathbb{E}\left[N_H(t)|N_A(t)=N_P(t)=0\right]$ is the expected size of the hypnozoite reservoir in an uninfected (no blood-stage infection) individual under $N$ rounds of MDA and is given by:

\begin{align}
   &\mathbb{E}\left[N_H(t)|N_A(t)=N_P(t)=0\right]
  \nonumber \\
   =&\begin{cases} 
   \int_0^t \frac{\nu p_H(t-\tau)\lambda(\tau)\big(1-e^{-\gamma (t-\tau)}\big) }{[1+\nu p_A(t-\tau)]^2 }d\tau & \text{if}\ t< s_1\\
   \int_{s_{n}}^t \frac{\nu p_H(t-\tau)\lambda(\tau)\big(1-e^{-\gamma (t-\tau)}\big) }{[1+\nu p_A(t-\tau)]^2 }d\tau \\ +
   \int_{0}^{s_1} \frac{\nu p_H^r(t-\tau,s_1-\tau)\lambda(\tau)\big(1-(1-p_{blood})e^{-\gamma (t-\tau)}\big) }{[1+\nu p_A^r(t-\tau,s_1-\tau)]^2 }d\tau\\+
   \int_{s_{1}}^{s_2} \frac{\nu p_H^r(t-\tau,s_1-\tau,s_2-\tau)\lambda(\tau)\big(1-(1-p_{blood})^2e^{-\gamma (t-\tau)}\big) }{[1+\nu p_A^r(t-\tau,s_1-\tau,s_2-\tau)]^2 }d\tau \\ +\ldots +
   \int_{s_{N-1}}^{s_N} \frac{\nu p_H^r(t-\tau,s_1-\tau,\ldots,s_n-\tau)\lambda(\tau)\big(1-(1-p_{blood})^Ne^{-\gamma (t-\tau)}\big) }{[1+\nu p_A^r(t-\tau,s_1-\tau,\ldots,s_N-\tau)]^2 }d\tau & \text{if}\ t\ge s_N
\end{cases}
\end{align}

The time-dependent parameters $p_1(t),\ p_2(t),\ k_1(t)$, and $k_T(t)$ that characterise the hypnozoite dynamics at the population level, account for all the infective bites received throughout time and change instantaneously with MDA because of the assumption of the instantaneous effect of the drug.

As these parameters involve numerical integration, we implement our own integro differential equation (IDE) solver using a $4^{th}$-order Runge-Kutta method, as described by Algorithm 1 in Anwar \textit{et al.} \cite{anwar2022multiscale}. Considering treatment at times $s_1,\ s_2,\ \ldots,\ s_N$, the parameters $p_1(t),\ p_2(t),\ k_1(t)$, and $k_T(t)$ are first obtained from the within-host model at each time $t$ to then obtain the solution of the population-level model at time $t$.  

\subsection{Optimisation model for the MDA intervals}
In this section, we construct a mathematical optimisation model to obtain the optimal timing for each MDA round. Suppose $s_1, s_2, \ldots, s_N$ are the $N$ MDA administration times. We want to optimise the MDA intervention times so that the outcome of the MDA implementation is optimised. We construct the optimisation problem as:

\begin{equation*}
\begin{aligned}
\minimise_{s_1, s_2,\ldots, s_N} \quad & \ Z\\
\textrm{s.t.} \quad & 0<s_1<s_2<\ldots<s_N,\\
\end{aligned}
\end{equation*}
where $Z$ is the objective function to be minimised. Based on good public health, we investigate two objective functions: 
\begin{itemize}
    \item $Z_1=\min_t \Big(I(t)+k_T(t)L(t)\Big),$
    \item $Z_2=\min_t \Big(\big(I(t)+L(t))W_h+(E_m(t)+I_m(t)\big)W_m$ \Big),
\end{itemize}
where $W_h$,\ $W_m$ are weighting factors for the human and mosquito population, respectively and $t\in [s_1\  t_{max}]$.
That is, $Z_1$ is the minimum of the sum of the blood-stage infected proportion and the average hypnozoite burden in liver-stage infected individuals for $t\in [s_1\  t_{max}]$ and $Z_2$ is the minimum of the weighted sum of the proportion of infected humans (both blood-stage and liver-stage) and infected mosquitoes (exposed and infectious) for $t\in [s_1\  t_{max}]$. Since the \textit{P. vivax} transmission is mainly dominated by hypnozoite dynamics, it is worth exploring the optimum effect of the drugs on disease prevalence and hypnozoite burden with the objective function, $Z_1$. As mosquito populations are an integral part in \textit{P. vivax} transmission, we explore the potential effect of infected (exposed and infectious) mosquitoes along with infected humans with the objective function $Z_2$. By setting $W_m=0$, we can also  investigate the optimal effect on only the human infected proportions (see Figure \ref{fig:opt_MDA2_MDA3}, for example).

\subsubsection{Without seasonality}
When seasonality is not considered, the time of the first MDA, $s_1$, can be considered arbitrary (as long as the dynamics have reached an equilibrium). In this case, we can fix $s_1=0$ (without loss of generality) and then the remaining MDA implementation times are optimised. Here, we minimise the objective function $Z$ over the time period of $[s_1\ t_{max}]$. We reconstruct the optimisation problem in terms of time intervals between MDA rounds. Let $x_1,\ x_2,\ \ldots,\ x_{N-1}$ be the time intervals between the first and second rounds of MDA, second and third rounds of MDA, and so on, respectively. That is, $x_1=s_2-s_1,\ x_2=s_3-s_2,\  \ldots, x_{N-1}=s_N-s_{N-1}$.
Then the optimisation problem becomes:

\begin{equation}
\begin{aligned}
\minimise_{x_1, x_2,\ldots, x_{N-1}} \quad & Z \\
\textrm{s.t.} \quad & x_1,\ x_2,\ \ldots,\ x_{N-1}>0\  \text{and}\ \sum x_i \le t_{max}\\
\end{aligned}
\label{eqn:optimum}
\end{equation}

\subsubsection{With seasonality}
When considering seasonality in the mosquito population, the time of the first MDA round is no longer arbitrary as the dynamics are periodic oscillations around the mean annual prevalence. As the periodic function that governs the mosquito birth rate, $b_m(t)$, has an oscillation  period of one year (assumed), the dynamics of human and mosquito populations have a peak within each year. Our optimisation problem without seasonality (Equation (\ref{eqn:optimum})) is constructed in terms of MDA intervals $x_1,\ x_2,\ \ldots,x_{N-1}$; for the case when we consider seasonality, we set a range of two years (starting from the time when prevalence is at a peak) for the optimisation algorithm to find the first MDA time, $s_1$. Here we define $x_0=s_1-\theta$ where $\theta$ is the peak prevalence time. That is, $x_0$ represents the interval between the prevalence peak time and the initial MDA time. The remaining times are obtained similarly without seasonality. Hence, the optimisation problem with seasonality in the mosquito population is:

\begin{equation}
\begin{aligned}
\minimise_{x_0, x_1,\ldots, x_{N-1}} \quad & Z\\
\textrm{s.t.} \quad &\ x_1,\ \ldots,\ x_{N-1}>0,\  x_0\ge0 \  \text{and}\ \sum x_i \le t_{max}\\
\end{aligned}
\label{eqn:optimum_ses}
\end{equation}

\section{Results}\label{result}
In this section, we present some numerical results. First, we consider the effect of MDA rounds if there were no seasonality. We explore the effect of one MDA round on disease prevalence (as a function of human to mosquito ratio, $m$), liver-stage infected proportions, and the hypnozoite reservoir in Section \ref{MDA1}. The effect of drug efficacy (varying $p_{rad}$) with one MDA round is presented in Section \ref{MDA1_efficacy}. We then present numerical results on the effect of multiple MDA rounds on disease prevalence (Section \ref{N_MDA}) by varying mosquito ratio where we present the rebound (e.g., minimum) disease prevalence obtained after 5 and 15 years for varying MDA rounds (up to $N=6$ rounds) with varying pre-MDA prevalence ($20\%-60\%$). Finally, we explore the effect of optimal MDA intervals on different disease prevalence  by varying mosquito ratios for the two different objective functions constructed in the previous section, both with and without seasonality (Section \ref{opt}).
\begin{figure}
    \begin{centering}
        \includegraphics[width=1\textwidth]{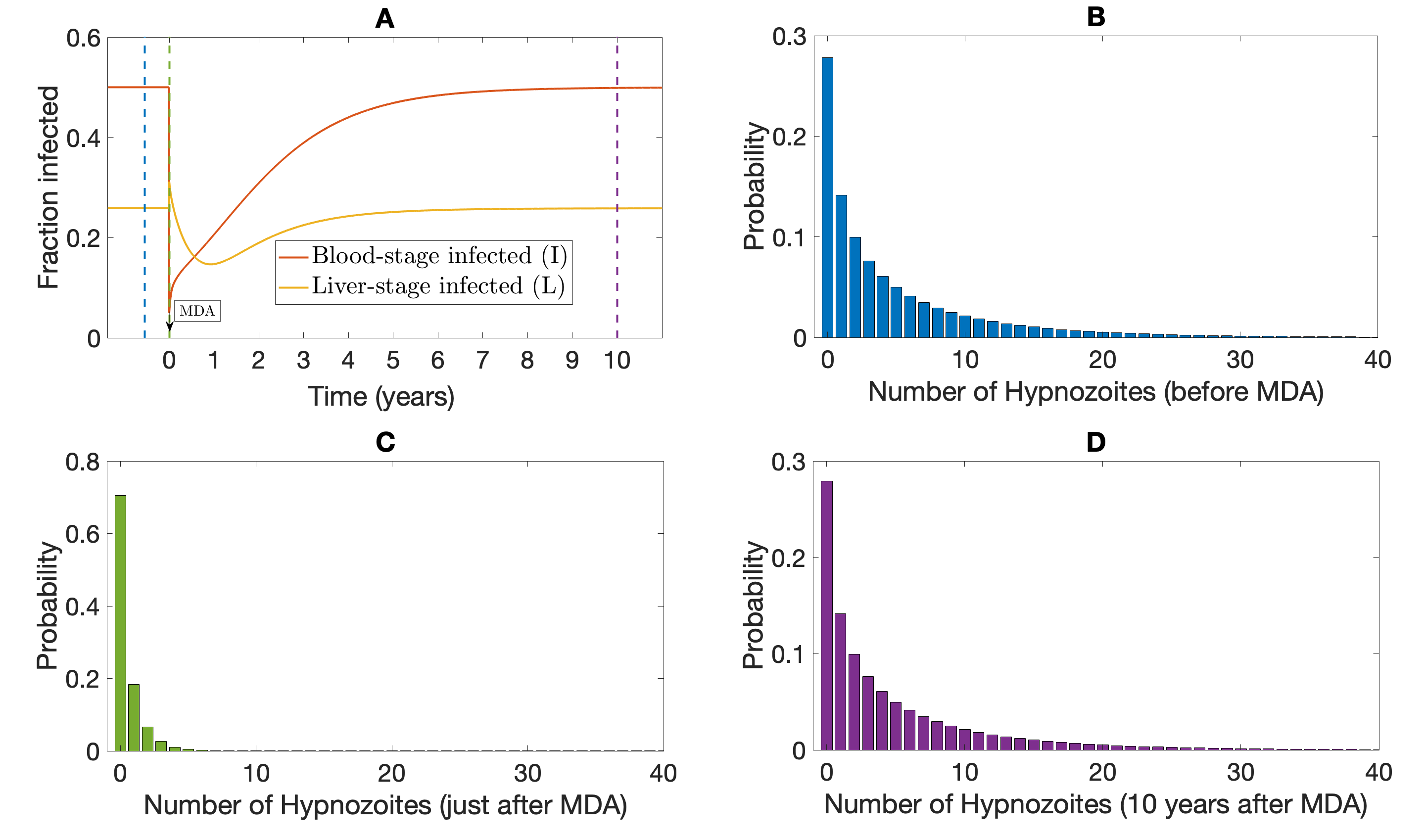}
    \caption{\textit{Results from multiscale model under radical cure treatment ($p_{blood}=0.9$, $p_{rad}=0.9$) with a single round of MDA without seasonality. Parameters are as per Table \ref{tab:white}. Subplot A depicts the proportion of blood-stage and liver-stage infected humans over time under treatment. The colored dashed lines indicate the times at which the hypnozoite distribution is quantified in Subplots B--D. Hypnozoite distribution in population (obtained as per Equations (74)--(75) in Mehra \textit{et al.} \cite{mehra2022hypnozoite}) before MDA (time is indicated by the blue dashed line in Subplot A is depicted in Subplot B. Subplot C and Subplot D depict the hypnozoite distribution in the population just after and 10 years after the MDA, respectively (times are indicated by green and purple dashed lines in Subplot A).}}
    \label{fig:bef_aft}
    \end{centering}
\end{figure}

\subsection{The effect of a single round of MDA (with \texorpdfstring{$p_{blood}=0.9,\  p_{rad}=0.9$}{})}\label{MDA1}

To quantify the effect of radical cure MDA, we first assume that one round of MDA is applied when the system is at a steady state (see Appendix \ref{SS} for detail on the steady-state derivation). Treatment coverage plays a significant role in the effect of an MDA program \cite{lydeamore2019biological}. To study the model behaviour, we assume that $100\%$ of the population is covered by the MDA scheme and that there is $90\%$ drug efficacy. Figure \ref{fig:bef_aft} shows the results from our multiscale model under one round of MDA. The drugs were assumed to have an instantaneous effect (with $p_{blood}=0.9$, $p_{rad}=0.9$); the hypnozoite reservoir size just before the MDA (Figure \ref{fig:bef_aft}B) becomes smaller in size (Figure \ref{fig:bef_aft}C; mode is $0$) as a result of the radical cure. That is, just immediately following MDA, most individuals will have no hypnozoites within their liver (with probability $\approx 0.7$). Disease prevalence drops significantly at the time of radical cure (Figure \ref{fig:bef_aft}A), as we assume that the drug clears any ongoing blood-stage infections with $90\%$ efficacy ($p_{blood}=0.9$). For liver-stage infected individuals, as the drugs are assumed to kill each hypnozoite with probability $p_{rad}=0.9$, the overall effect of the drug depends on the size of the hypnozoite reservoir. If the size of the hypnozoite reservoir is substantial before the treatment, the overall effect would be insignificant, and vice versa. As individuals are still exposed to infectious mosquito bites, and each infective bite contributes to an average of $\nu$ number of hypnozoites that activate at a constant rate $\alpha$, both blood-stage and liver-stage proportions reach the same equilibrium state (Figure \ref{fig:bef_aft}D) as before MDA (Figure \ref{fig:bef_aft}B) eventually.\par

\begin{figure}
    \begin{centering}
        \includegraphics[width=\textwidth]{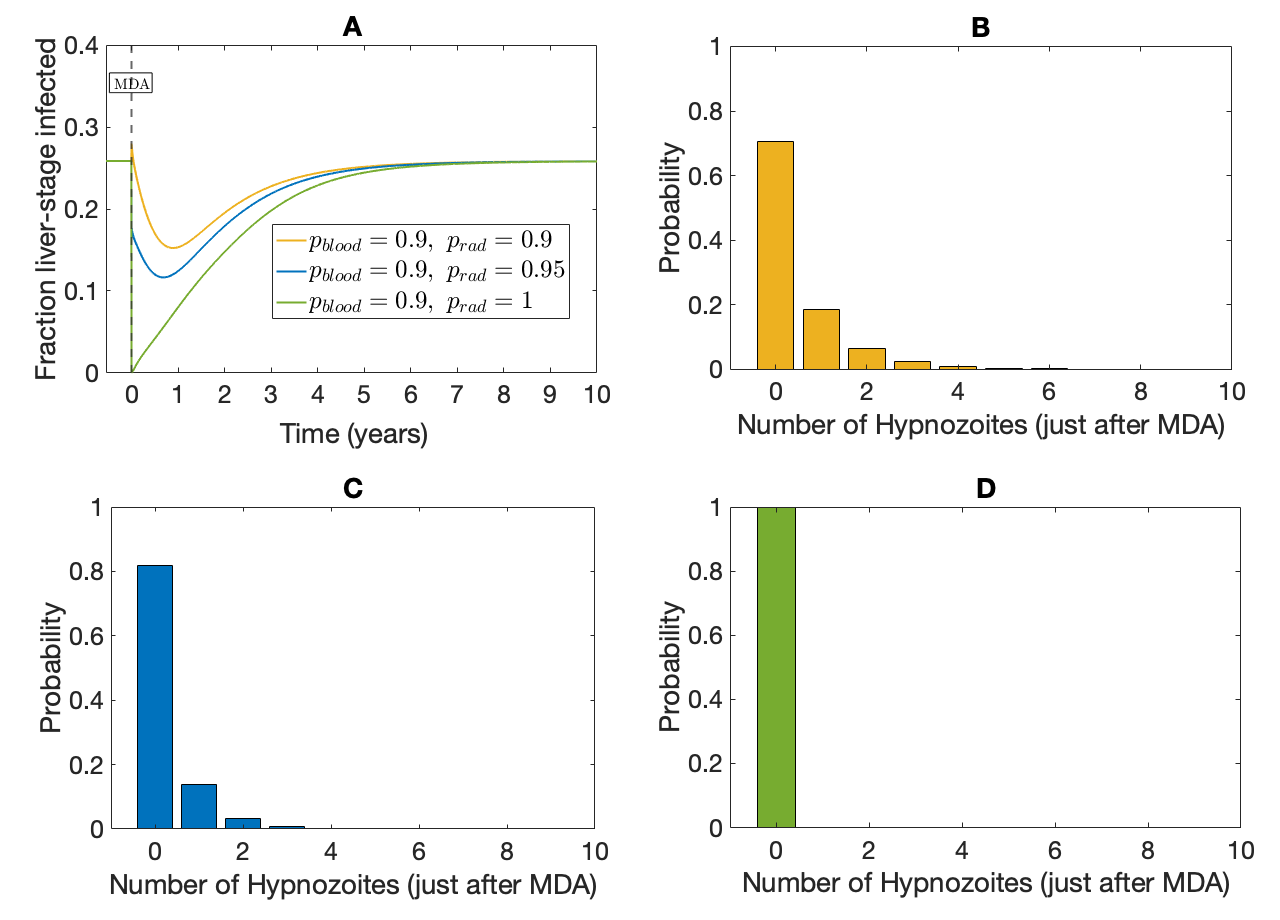}
    \caption{\textit{Effect of radical cure on liver-stage infected individuals without seasonality. Subplot A depicts the proportion of liver-stage infected for different hypnozoitocidal efficacy levels ($p_{rad}$). Yellow, blue, and green lines corresponds to $p_{rad}=0.9$, $p_{rad}=0.95$, and $p_{rad}=1$, respectively. Here $p_{blood}=0.9$ for all scenarios. Subplots B, C, and D show the hypnozoite distribution within the population just after the MDA program when $p_{rad}=0.95$, $p_{rad}=0.99$, and $p_{rad}=1$, respectively (obtained as per Equations (74)$\mbox{--}$(75) in Mehra \textit{et al.} \cite{mehra2022hypnozoite}). Other parameters are as in Table \ref{tab:white}.}}
    \label{fig:H_dif_efficacy}
    \end{centering}
\end{figure}

\begin{figure}[ht]
    \begin{centering}
        \includegraphics[width=1\textwidth]{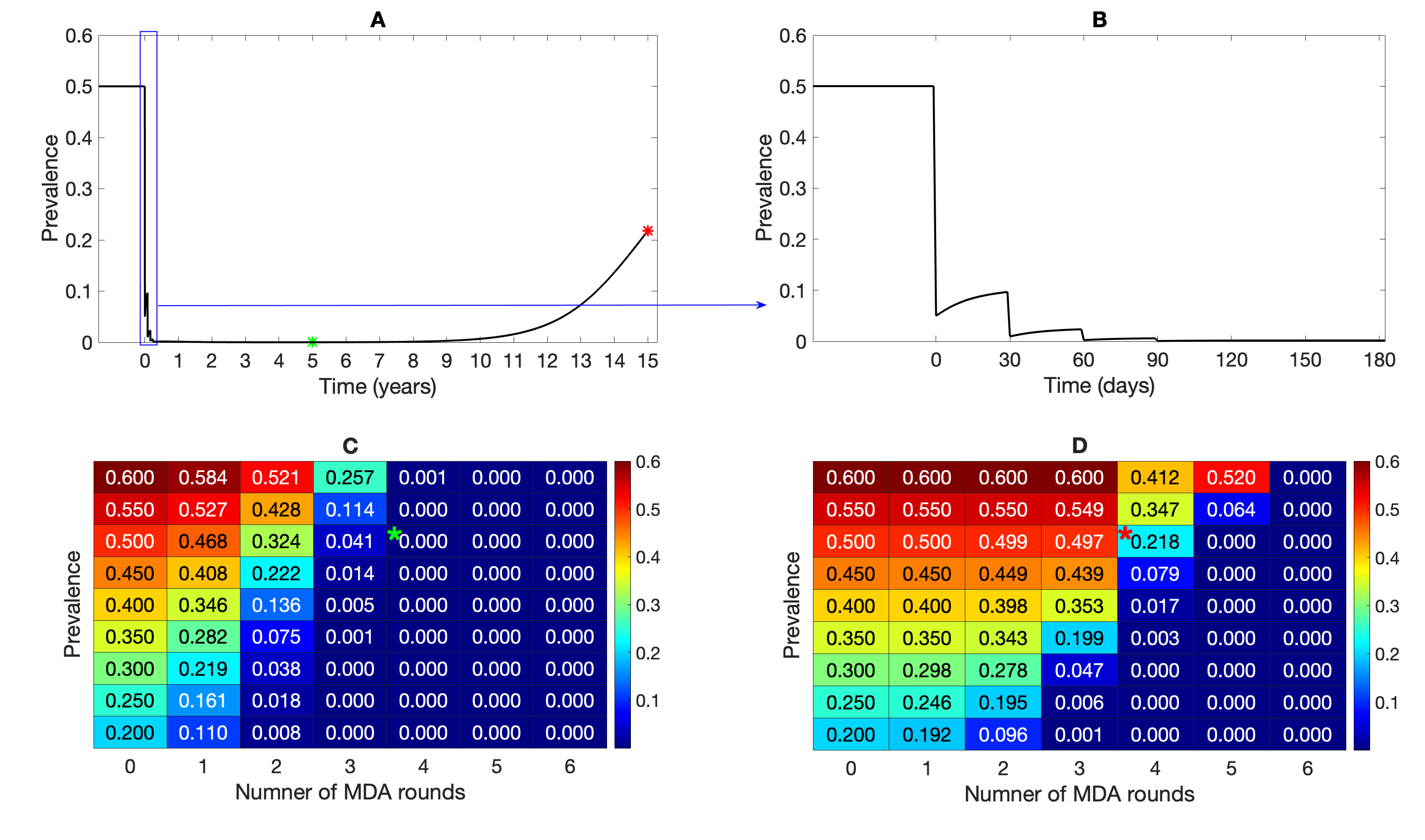}
    \caption{\textit{Effect of multiple rounds of MDA without seasonality. Subplot A shows the effect of four MDA rounds on prevalence over time whereas Subplot B is a snapshot from Subplot A that shows the transient dynamics during the MDA rounds. A sensitivity analysis of up to $N=6$ MDA rounds over different steady state prevalences is illustrated in Subplots C$\mbox{--}$D. Subplot C shows the disease prevalence 5 years after the first MDA round for up to six MDA rounds and  Subplot D shows the disease prevalence 15 years after the first MDA round. The green and red asterisks in Subplots C and D are the prevalences corresponding to the green and red asterisks in Subplot A, respectively. The intervals between MDA rounds are fixed at 30 days. Other parameters are as in Table \ref{tab:white}}.}
    \label{fig:MDA6_sen}
    \end{centering}
\end{figure}

\subsection{The effect of a single round of MDA, varying \texorpdfstring{$p_{blood}$}{}}\label{MDA1_efficacy}

The effect of the drug on disease transmission and the hypnozoite reservoir also changes with the efficacy of the drug (Figure \ref{fig:H_dif_efficacy}). Figure \ref{fig:H_dif_efficacy}A illustrates the effect of varying efficacies of the hypnozoicidal drug (i.e., $p_{rad}$) on liver-stage infected proportions. Figure \ref{fig:H_dif_efficacy}B illustrates the hypnozoite distribution just after the application of MDA when $p_{blood}=p_{rad}=0.9$ and Figure \ref{fig:H_dif_efficacy}C illustrates the hypnozoite distribution when $p_{blood}=0.9,\ p_{rad}=0.95$. If the hypnozoicidal drug were 100\% effective (that is, $p_{rad}=1$) then all liver-stage infected individuals would recover (green line in Figure \ref{fig:H_dif_efficacy}A). In the case of $p_{rad}=1$, the hypnozoite reservoir within the human population would be completely cleared (Figure \ref{fig:H_dif_efficacy}D). In other words, immediately following drug administration, no individuals would be liver-stage infected. However, the disease will eventually reach the same equilibrium state as if no treatment were administered.  (Figure \ref{fig:H_dif_efficacy}A).\par

\subsection{The effect of multiple MDA rounds}\label{N_MDA}

We also examined the impact of multiple MDA rounds on transmission and hypnozoite dynamics in the absence of seasonality (Figure \ref{fig:MDA6_sen}). Figure \ref{fig:MDA6_sen}A depicts the long-term behaviour of the transmission dynamics under four MDA rounds where the transient behaviour over the time of the MDA rounds (150 days) is depicted in Figure \ref{fig:MDA6_sen}B. Here, we assumed a fixed interval (30 days) between MDA rounds, although intervals between MDA rounds among studies vary widely, from weeks to several months \cite{newby2015review}. The effect of four successive MDA rounds is clearly visible in Figure \ref{fig:MDA6_sen}A. Disease prevalence was driven down to approximately zero after the fourth round. However, as we model the system as a deterministic process and the effect of the drug is temporary, over time the disease reaches the same endemic steady state as before treatment. The overall effect of radical cure MDA treatment also depends on the disease prevalence; the lower the prevalence, the more effective the MDA in reducing the disease prevalence and hypnozoite-positive proportions. Figure \ref{fig:MDA6_sen}C and D illustrate the sensitivity analysis of up to six MDA rounds at different assumed prevalences ($20\%\mbox{--}60\%$) obtained by varying mosquito ratio, $m$, showing the rebound prevalence 5 years and 15 years after the first MDA round was applied. The interval between each MDA round was again fixed at 30 days. If the prevalence before MDA is high, the dynamics reach the equilibrium state faster than when the prevalence is low before MDA.

\subsection{Optimal MDA programs}\label{opt}
\subsubsection{Without seasonality}

\begin{figure}[ht]
    \begin{centering}
\includegraphics[width=\textwidth]{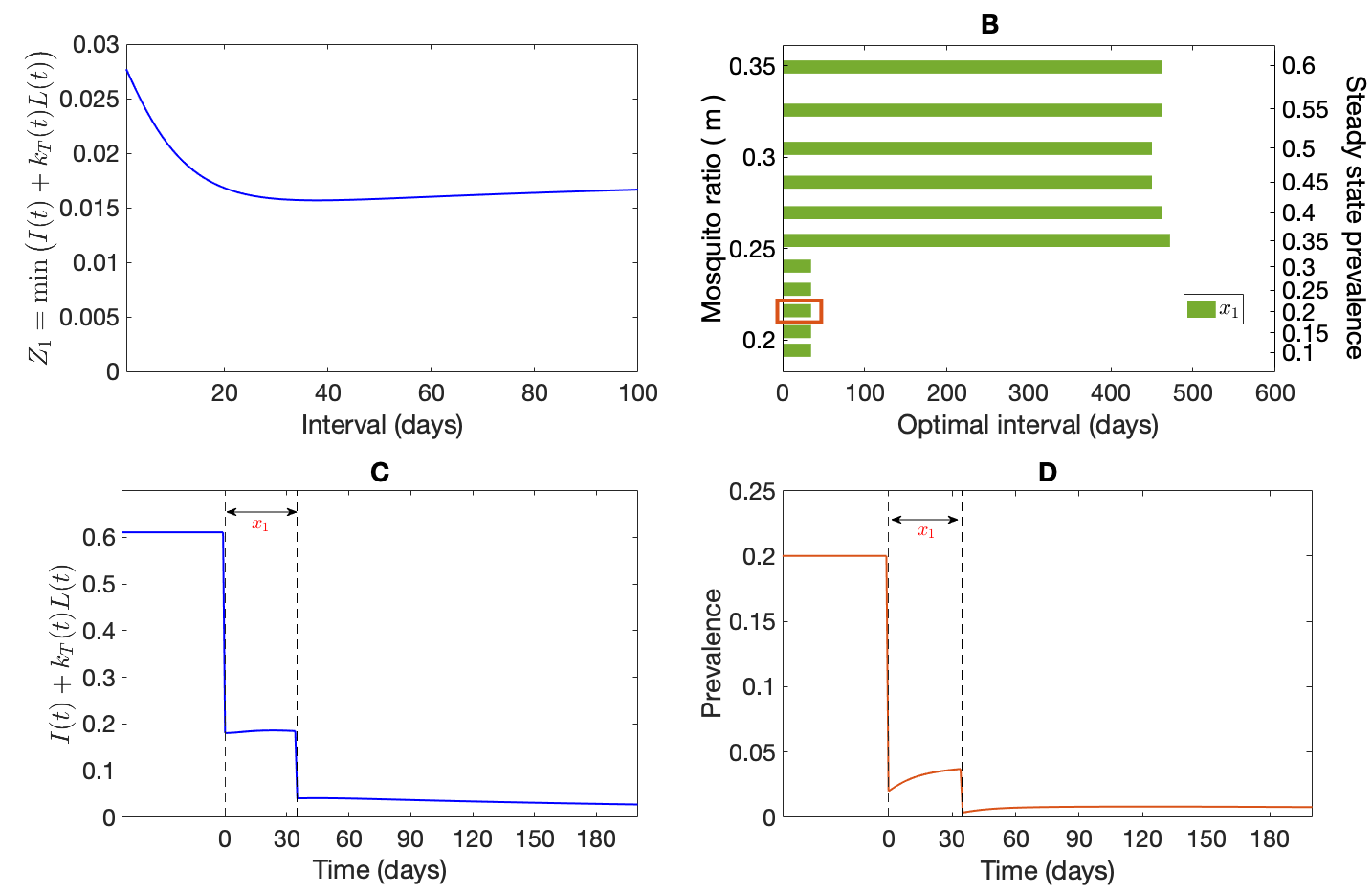}
  \caption{\textit{Effect of two rounds of optimally timed MDA. Subplot A shows the impact of varying intervals between two MDAs on the objective function $Z_1$ (using a starting steady state disease prevalence of 20\%) where the minimum objective value obtained from the optimisation problem (Equation (\ref{eqn:optimum})) is when the interval is $34.7$ days. Subplot C depicts the objective function, $Z_1$, over time, before and after the MDAs when using the optimal interval of $\approx 35$ days.  Subplot D illustrates the transient disease dynamics corresponding to two optimally timed MDAs (time for the first MDA is arbitrary), separated in time by $\approx 35$ days. Finally, subplot B illustrates the optimal interval for different disease prevalences (right vertical axis) by varying the mosquito ratio (left vertical axis), $m$ corresponding to the objective function $Z_1$. These optimal intervals are for two MDA rounds found by solving Equation (\ref{eqn:optimum}) where the red rectangle shows the $\approx 35$ day optimal interval for the 20\% steady state disease prevalence used in Subplot A, C, and D. All parameters are as in Table \ref{tab:white}.}}
    \label{fig:opt_MDA2_Z1}
    \end{centering}
\end{figure}

\begin{figure}[htbp]
    \begin{centering}
        \includegraphics[width=\textwidth]{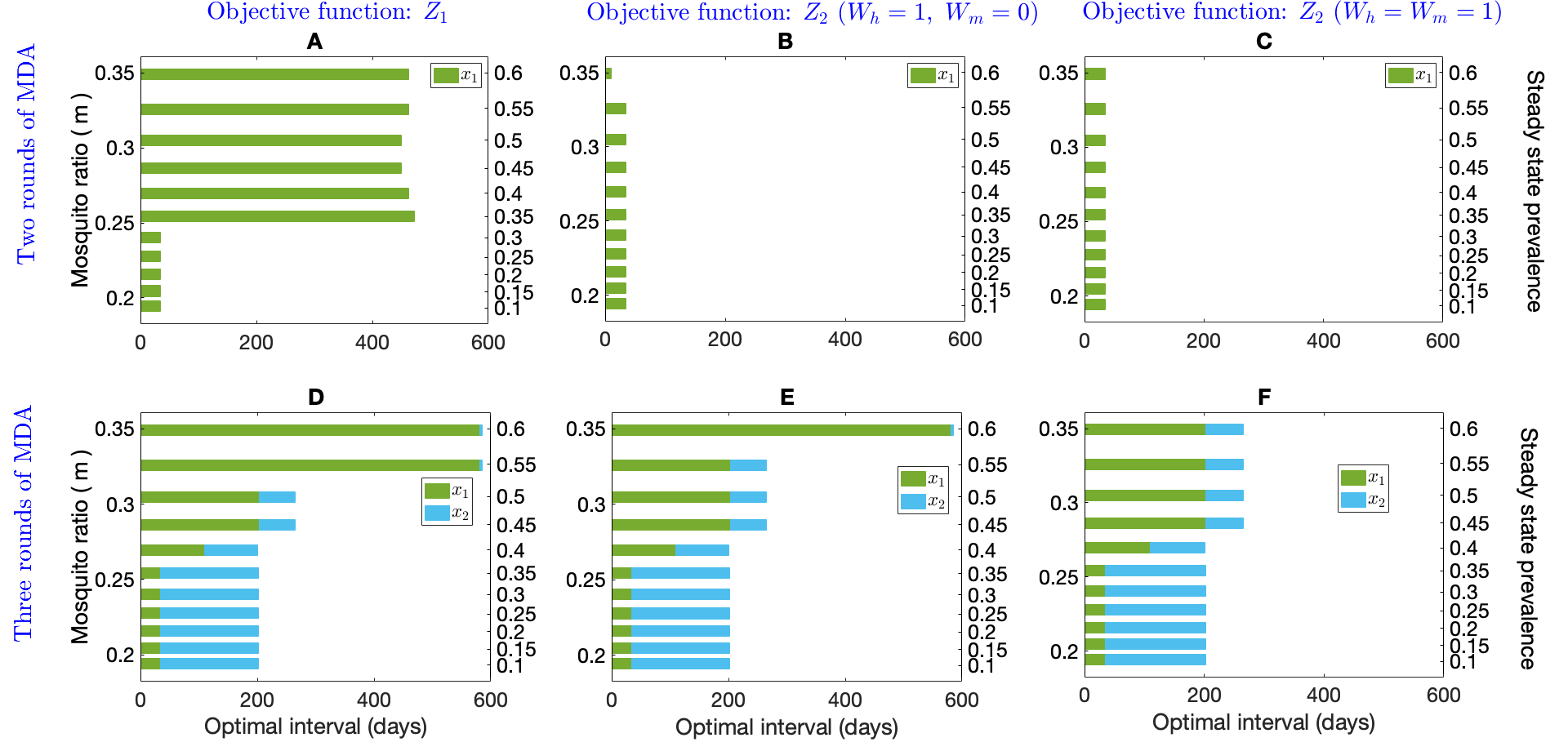}
    \caption{\textit{Sensitivity analysis of two and three rounds of optimal MDA intervals over different disease prevalence (right vertical axis) without seasonality. Prevalence is varied by varying mosquito ratio, $m$ (left vertical axis).  The optimal interval between two and three rounds of MDA for the objective function $Z_1$ (Subplot A, D respectively), $Z_2$ with $W_h=1,\ W_m=0$ (Subplot B, E respectively) and $Z_2$ with $W_h=1,\ W_m=1$ (Subplot C, F respectively) are shown. All parameters are as in Table \ref{tab:white}.}}
    \label{fig:opt_MDA2_MDA3}
    \end{centering}
\end{figure}

To obtain the optimal interval between MDA rounds, we use the optimisation problem defined in Equation (\ref{eqn:optimum}). We used the MATLAB optimisation tool `Multistart' (with 80 different initial starting points) with \verb|fmincon| (SQP algorithm) to generate global optimal solutions. The results of two optimally timed MDA rounds are illustrated in Figure \ref{fig:opt_MDA2_Z1} for a steady state disease prevalence (see Appendix \ref{SS} for the derivation of the steady state disease prevalence) of 20\% with the objective function $Z_1$. With our choice of parameter values (see Table \ref{tab:white}), the optimisation problem gives an optimal interval of $x_1=s_2-s_1=34.7$ days, as illustrated in Figure \ref{fig:opt_MDA2_Z1}A. Figure \ref{fig:opt_MDA2_Z1}C depicts the sum of blood-stage infected population proportion and the hypnozoite burden on liver-stage infected population over time, $I(t)+k_T(t)L(t)$, before and after the MDA rounds using the optimal interval of $x_1=34.7$ days. The effect of the optimally timed MDA rounds on disease prevalence (20\%) is depicted in Figure \ref{fig:opt_MDA2_Z1}D. The dashed vertical lines in Figure \ref{fig:opt_MDA2_Z1}C and D indicate the optimal time for the MDA rounds ($x_1=34.7$). When no seasonality is considered, the time of the first MDA can be at any arbitrary time (after an equilibrium has been reached). The equilibrium disease prevalence (obtained by varying mosquito ratio, $m$) greatly affects the optimum intervals (Figure \ref{fig:opt_MDA2_Z1}B). The left vertical axis in Figure \ref{fig:opt_MDA2_Z1}B illustrates the mosquito ratio, $m$, and the values on the right vertical axis depict the prevalence corresponding to each green bar. For higher prevalence ($25\%\mbox{--}60\%$), the optimisation problem with the objective function $Z_1$ suggests an interval of around 480 days between the two MDA rounds.\par

Figure \ref{fig:opt_MDA2_MDA3} shows the optimum interval for two (first row) and three (second row) MDA rounds for different equilibrium disease prevalences (right vertical axis) obtained through changing the  mosquito ratio, $m$, with three different choices of the objective function. The first, second, and third columns represent the objective function $Z_1$, $Z_2$ with $W_h=1,\ W_m=0$, and $Z_2$ with $W_h=1,\ W_m=1$, respectively. In contrast with the objective function $Z_2$ with $W_h=1,\ W_m=0$ and $Z_2$ with $W_h= W_m=1$, the optimisation problem suggests a longer interval for the second MDA round for higher prevalences ($>35\%$) with $Z_1$ when only two rounds of MDA are used (Figure \ref{fig:opt_MDA2_MDA3}B). The interval between the two MDA rounds is very similar for different prevalences for the objective function $Z_2$ with $W_h=1\ W_m=0$ and $Z_2$ with $W_h= W_m=1$ (Figure \ref{fig:opt_MDA2_MDA3}B, C).\par

 The optimal intervals for three MDA rounds depend on both $m$, hence prevalence, and the choice of the objective function (Figure \ref{fig:opt_MDA2_MDA3}D-F). If three MDA rounds are considered, the optimisation problem (Equation (\ref{eqn:optimum})) suggests a similar interval for all of the MDA rounds with all three choices of the objective function for low prevalence ($<50\%$). 
 But for higher prevalences ($>55\%$), for $Z_1$ and $Z_2$ with $W_h=1,\ W_m=0$, the optimisation routine suggests an immediate implementation of the third round of MDA after a long delay in between. However, for $Z_2$ with $W_h=1,\ W_m=0$, the interval $x_2$ becomes shorter as $m$, hence prevalence gets higher (but remains the same).\par

\subsubsection{With seasonality}

\begin{figure}[htbp]
    \begin{centering}
        \includegraphics[width=\textwidth,scale=2.5]{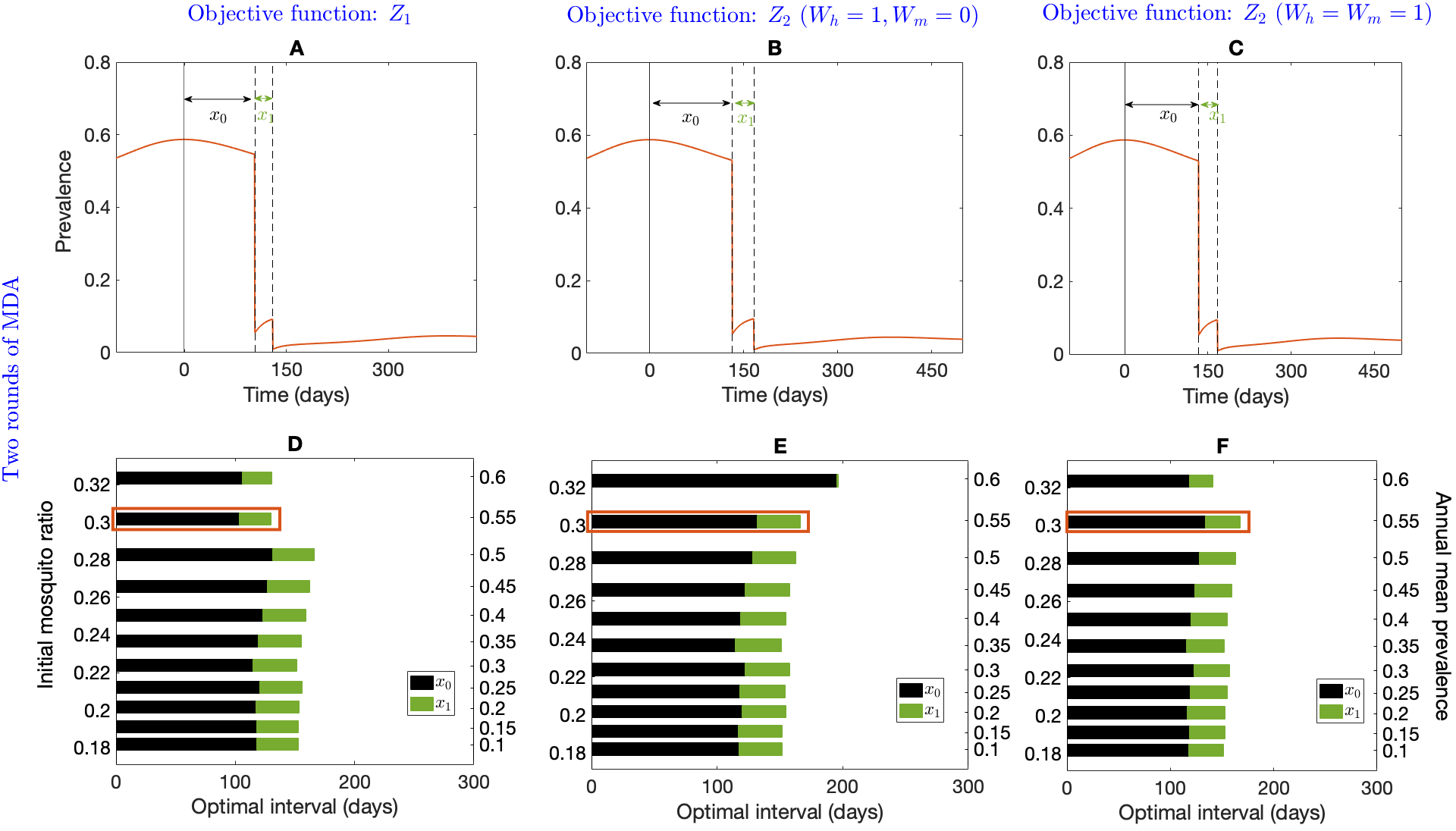}
    \caption{\textit{Effect of two rounds of optimally timed MDA with mosquito seasonality. Subplots A-C depict the impact of optimal MDA on disease prevalence (annual mean disease prevalence before MDA of 54.9\%) with objective function $Z_1$, $Z_2\, (W_h=1,\ W_m=0)$, and $Z_2\, (W_h=W_m=1)$, respectively. The solid vertical line indicates the time when the prevalence reaches a peak before the initial MDA. The dashed vertical lines indicate the optimal times for the MDA rounds. Subplots D$\mbox{--}$F depict the sensitivity analysis over different annual mean disease prevalences with the objective function $Z_1$, the objective function $Z_2$ with $W_h=1, W_m=0$, and objective function $Z_2$ with $W_h=W_m=1$ respectively. All parameters are as in Table \ref{tab:white}.}}
    \label{fig:S_MDA2_all}
    \end{centering}
\end{figure}

The effect of two optimally timed MDA rounds (including the first round, which was not required to be considered when seasonality was not considered) is illustrated in Figure \ref{fig:S_MDA2_all}. The optimal time for the first MDA round is approximately the same for different annual mean disease prevalences (right vertical axis, obtained by varying initial mosquito ratio, $m_0$) and the objective functions (Figure \ref{fig:S_MDA2_all}D\mbox{--}F). The seasonal amplitude, $\eta$, is thought to play an important role in intervention strategies \cite{selvaraj2018seasonality}; here we have assumed $\eta=0.1$. Figure \ref{fig:S_MDA2_all}A\mbox{--}C shows the impact of two MDA rounds on disease prevalence for all the objective functions when there is a 54.9\% annual mean disease prevalence before MDA for demonstrative purposes. The vertical solid line indicates the time when the pre-MDA prevalence reaches a peak and the vertical dashed lines indicate the time of the MDA implementations. When the annual mean disease prevalence is 54.9\%, the optimisation problem with our choice of parameters as per Table \ref{tab:white} along with the objective function $Z_1$, provides the interval $x_0=103.4$ days and $x_1=26.7$ days for two MDA rounds. The intervals with $Z_2\, (W_h=1,W_m=0)$ are $x_0=132.1$ days, $x_1=34.3$ days and with $Z_2\, (W_h=W_m=1)$ are $x_0=133.7$ days and $x_1=33.7$ days.  The sensitivity analysis for optimal interval time with different annual mean prevalences (right vertical axis, corresponding to each bar) is illustrated in Figure \ref{fig:S_MDA2_all}D\mbox{--}F with $Z_1$, $Z_2\, (W_h=1,\ W_m=0)$ and $Z_2\, (W_h=W_m=1)$, respectively. The red rectangles in Figure \ref{fig:S_MDA2_all}D\mbox{--}F indicate the optimal intervals corresponding to Figure \ref{fig:S_MDA2_all}A\mbox{--}C. With respect to all objective functions, $Z_1$, $Z_2\, (W_h=1,\ W_m=0)$, and $Z_2\, (W_h=W_m=1)$, the optimal intervals are very similar when the mean annual prevalence is low ($<50\%$). However, the optimisation algorithm suggests an immediate implementation for the two MDA rounds for higher annual mean prevalence with the objective function $Z_2\, (W_h=1,\ W_m=0)$ (Figure \ref{fig:S_MDA2_all}E), while with $Z_2\, (W_h=W_m=1)$, the algorithm suggests a similar interval as for low prevalences (Figure \ref{fig:S_MDA2_all}F). With objective function $Z_1$, the interval between the two MDA rounds is also very similar for different annual mean prevalences (Figure \ref{fig:S_MDA2_all}D).  \par

\begin{figure}[htbp]
    \begin{centering}
        \includegraphics[width=\textwidth,scale=3.5]{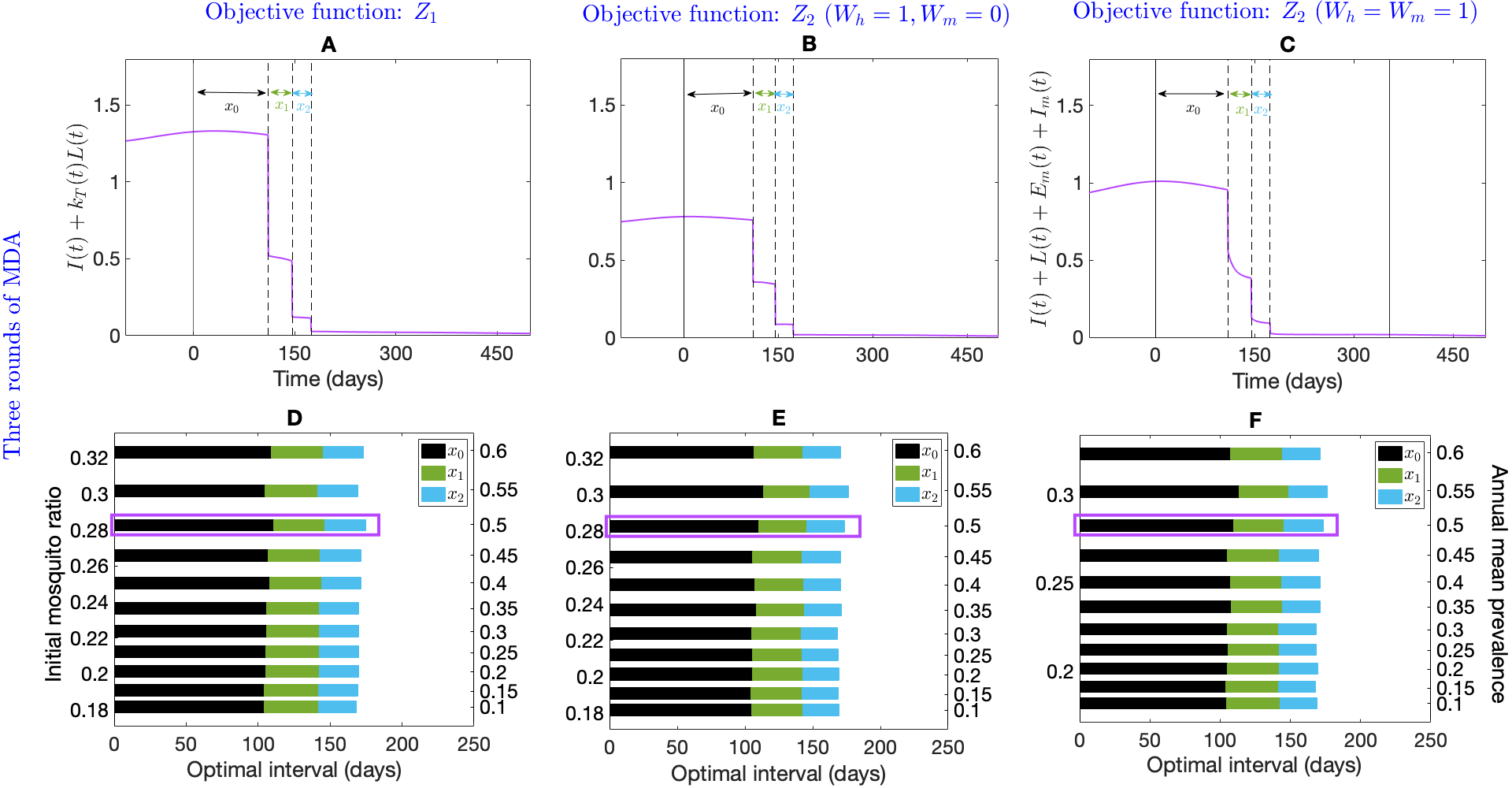}
    \caption{\textit{Effect of three rounds of optimally timed MDA with mosquito seasonality. Subplots A$\mbox{--}$C depict the impact of three optimally timed MDAs on objective functions $Z_1$, $Z_2\,(W_h=1,\ W_m=0)$, and $Z_2\, (W_h=W_m=1)$, respectively (note the change in X-axis between Subplots A and B$\mbox{--}$C) (annual mean disease prevalence before MDA of 49.9\%). Subplots D-F depict a sensitivity analysis over different annual mean disease prevalences (right vertical axis) obtained by varying initial mosquito ratio (left vertical axis) with objective function $Z_1$, the objective function $Z_2$ with $W_h=1, W_m=0$, and the objective function $Z_2$ with $W_h=W_m=1$, respectively. All parameters are as in Table \ref{tab:white}.}}
    \label{fig:S_MDA3_all}
    \end{centering}
\end{figure}

Figure \ref{fig:S_MDA3_all} shows the optimal intervals when three MDA rounds are considered for each objective function. Figure \ref{fig:S_MDA3_all}A\mbox{--}C demonstrates the effect of three optimally timed MDA rounds on the objective function $Z_1$, $Z_2\, (W_h=1,\ W_m=0)$ and $Z_2\, (W_h=W_m=1)$, respectively over time where the  vertical solid line indicates the time when the prevalence reaches a peak and the three subsequent vertical dashed lines indicate the optimal time for the three MDA rounds. The sensitivity analysis for optimal interval time with different annual mean prevalences (right vertical axis) is illustrated in Figure \ref{fig:S_MDA3_all}D\mbox{--}F with $Z_1$, $Z_2\, (W_h=1,\ W_m=0)$ and $Z_2\, (W_h=W_m=1)$, respectively where the violet rectangles in Figure \ref{fig:S_MDA3_all}D\mbox{--}F indicate the optimal intervals corresponding to Figure \ref{fig:S_MDA3_all}A\mbox{--}C. With respect to all objective functions, $Z_1$, $Z_2\, (W_h=1,\ W_m=0)$, and $Z_2\, (W_h=W_m=2)$, the optimal timing for the second and third MDA round, that is the interval between the last two rounds is almost identical throughout all different prevalences. \par


\begin{figure}[htbp]
    \begin{centering}
\includegraphics[width=\textwidth]{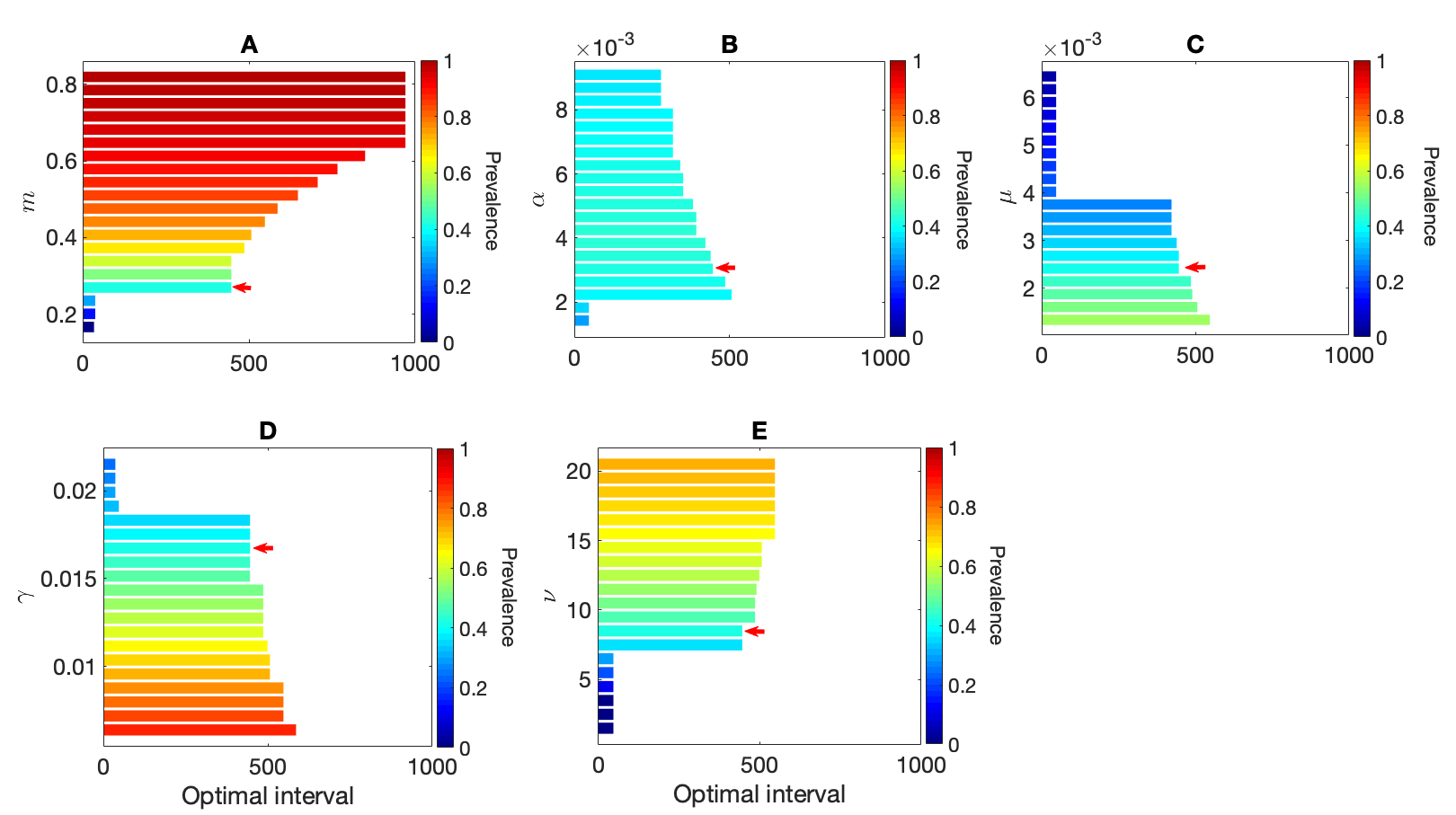}
  \caption{\textit{Effect of change in model parameters on the optimal interval (without seasonality) between two MDA rounds with the objective function $Z_1$ which is the minimum of the sum of the blood-stage infected proportion and the average
hypnozoite burden in liver-stage infected individuals at time $t$. Subplots A$\mbox{--}$E illustrate the impact of varying $m$ (number of mosquitoes per human), 
 $\alpha$ (hypnozoite
activation rate), $\mu$ (hypnozoite death rate), 
 $\gamma$ (natural recovery rate), and $\nu$ (average number of hypnozoites per bite) on the optimal interval, respectively. The colorbars in each subplot illustrate the equilibrium prevalence corresponding to the parameters before the first MDA was implemented. The red arrows in Subplots A$\mbox{--}$E indicate the baseline parameters in Table \ref{tab:white} and optimal interval when prevalence is around $40\%$ (see Figure \ref{fig:opt_MDA2_Z1}B as a reference). Parameter ranges for Subplot A$\mbox{--}$E are as in Table \ref{tab:white}.}}
    \label{fig:gbl_sen}
    \end{centering}
\end{figure}

\begin{figure}[htbp]
    \begin{centering}
\includegraphics[width=\textwidth]{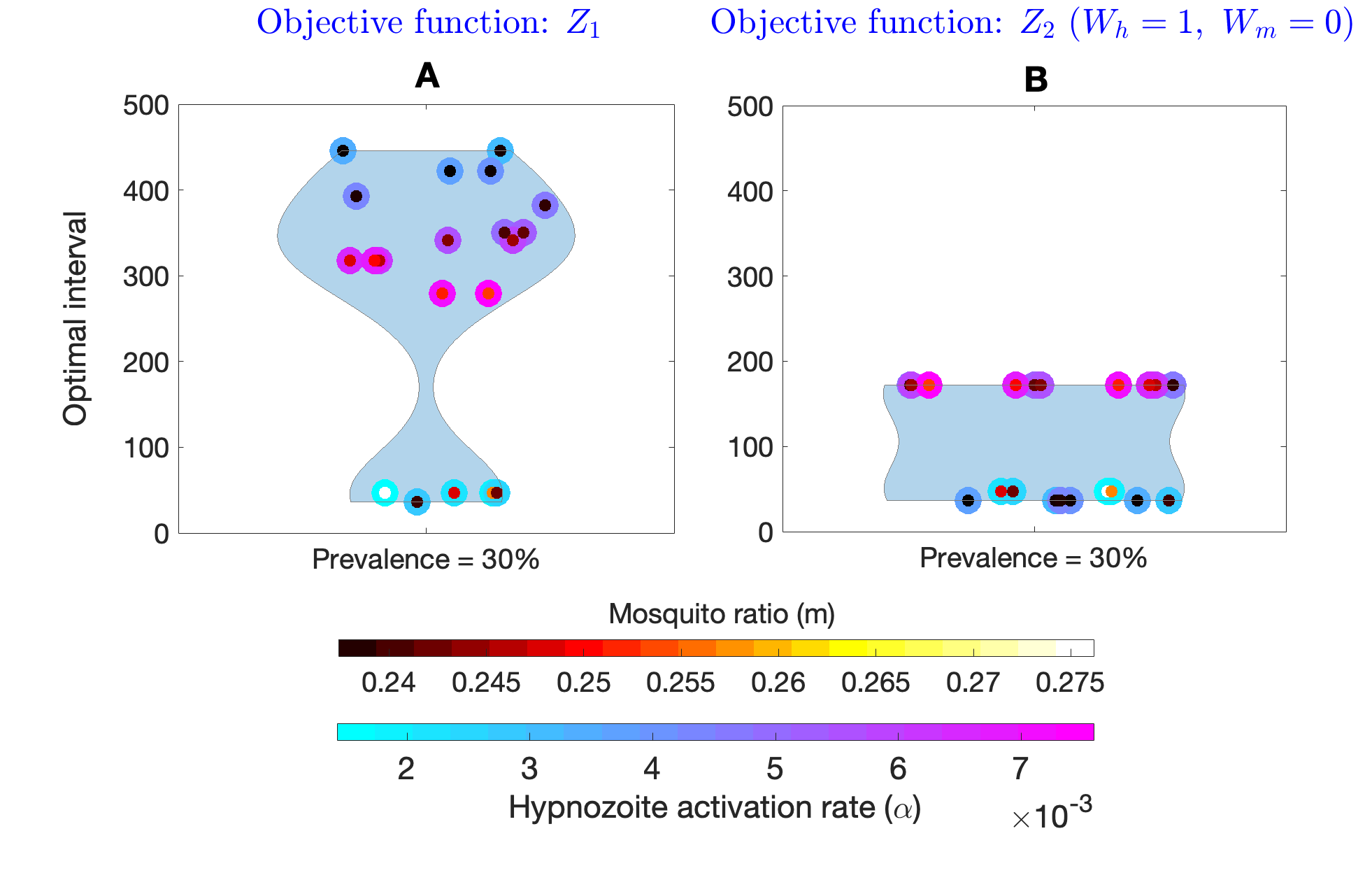}
  \caption{\textit{Effect of change in two model parameters (the hypnozoite activation rate, $\alpha$, 
and the number of mosquitoes per human, $m$) on the optimal interval (without seasonality) between two MDA rounds with the objective functions $Z_1$ and $Z_2$ with $W_h=1, W_m=0$.  Subplot A  is a violin plot that illustrates the optimal interval corresponding to the objective function $Z_1$, whereas Subplot B illustrates the optimal interval corresponding to the objective function $Z_2$ with $W_h=1, W_m=0$. In both cases, for a given value of $\alpha$ (outer color of the scatter points), we choose the parameter value $m$ (inner color of the scatter points) so that the steady-state prevalence is $30\%$. The colorbars on the bottom illustrate the value of $m$ and $\alpha$ respectively. All other parameters are as in Table \ref{tab:white}.}}
    \label{fig:violin}
    \end{centering}
\end{figure}

The choice of model parameters can significantly influence the optimal MDA intervals.  In order to obtain an equilibrium disease prevalence (without seasonality, see Appendix \ref{SS}) for Figures \ref{fig:MDA6_sen} and \ref{fig:opt_MDA2_Z1}, we only varied the human-to-mosquito ratio ($m$) and kept all other parameter values as per Table \ref{tab:white}. We note that there are (possibly) many other combinations of model parameters that could generate the same equilibrium prevalence (see Figure \ref{fig:gbl_sen}). Hence, we performed a sensitivity analysis for the parameters $m,\ \alpha,\ \mu,\ \gamma$, and $\nu$ on the optimal interval (without seasonality) for two MDA rounds with the objective function $Z_1$ which is the minimum of the sum of the blood-stage infected proportion and the average hypnozoite burden in liver-stage infected individuals at time $t$.  Figure \ref{fig:gbl_sen}A$\mbox{--}$E depict the effect of varying $m,\ \alpha,\ \mu,\ \gamma$, and $\nu$ on the optimal intervals, respectively. The color of the bars in Figure \ref{fig:gbl_sen}A$\mbox{--}$E depicts the equilibrium prevalence corresponding to the parameter value. The red arrows in Figure \ref{fig:gbl_sen}A$\mbox{--}$E depict the baseline parameters in Table \ref{tab:white} that generate a prevalence of $40\%$ as shown in Figure \ref{fig:opt_MDA2_Z1}B. As illustrated in Figure \ref{fig:gbl_sen}A, the abundance of mosquitoes can drastically influence the disease equilibrium as the force of reinfection (that is, the probability of reinfection per unit time) increases with $m$ $(\lambda= m_0abI_m)$. We see that the optimal intervals can be different for the same equilibrium prevalence generated with a different combination of the parameters $m,\ \alpha,\ \mu,\ \gamma$, and $\nu$. That is, the optimal interval without seasonality depends on the input model parameters. To investigate this further, we vary the values of $\alpha$\ and set a value of $m$ such that the steady-state disease prevalence is $30\%$ (Figure \ref{fig:violin}). Figure \ref{fig:violin}A illustrates the distribution of the optimal interval for two MDA rounds for different values of $\alpha$ and $m$. The optimal interval varies from around 47 days to 446 days for the different combinations of $\alpha$ and $m$ (objective function $z_1$). Figure \ref{fig:violin}B depicts the distribution of optimal intervals for the same set of parameters but with the objective function $Z_2$ with $W_h=1, W_m=0$. In this case, the optimal interval varies from around 37 days to 172 days. The results illustrate that prevalence alone is not sufficient to determine an optimal MDA interval when there is no seasonality in the mosquito population. \par

\begin{figure}[htbp]
    \begin{centering}
\includegraphics[width=\textwidth]{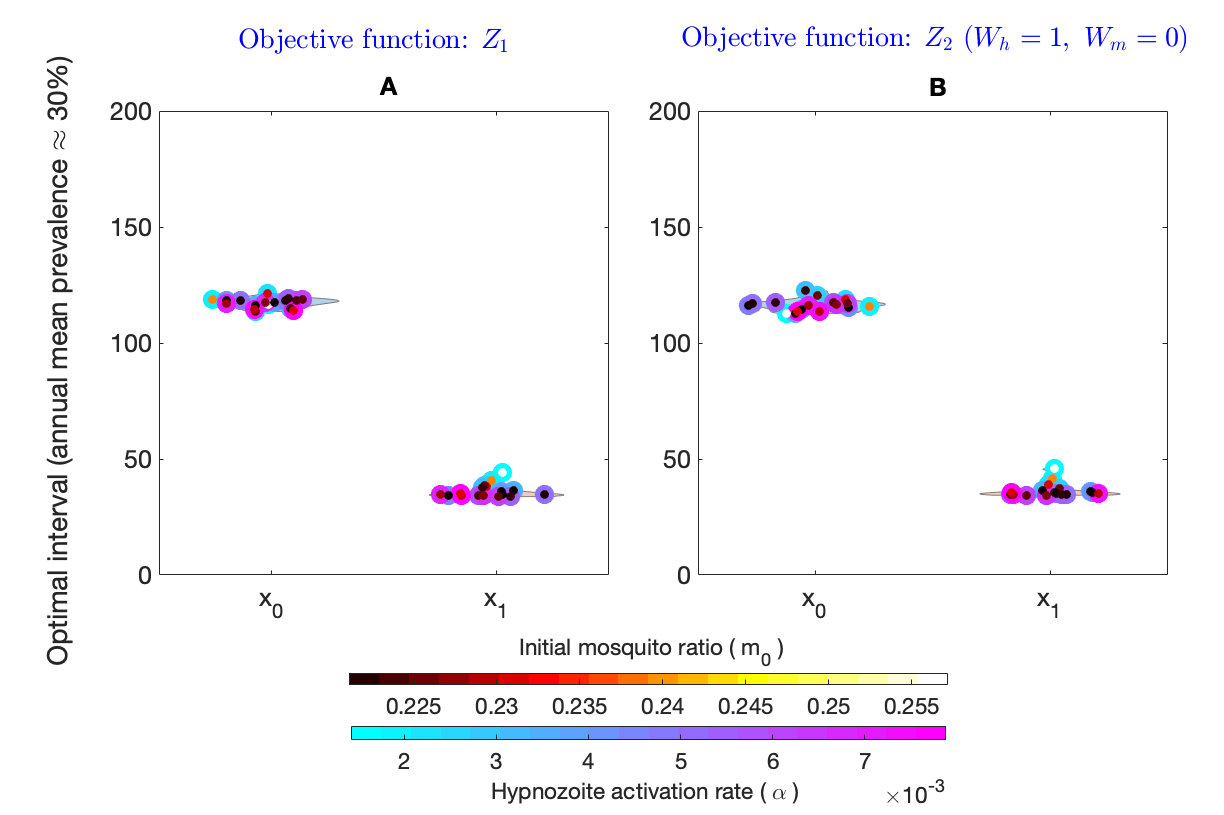}
  \caption{\textit{Effect of change in two model parameters (the hypnozoite activation rate, $\alpha$, 
and the initial number of mosquitoes per human, $m_0$) on the optimal interval (with seasonality) between two MDA rounds with the objective functions $Z_1$ and $Z_2$ with $W_h=1, W_m=0$.  Subplot A illustrates the optimal intervals $x_0$ and $x_1$ corresponding to the objective function $Z_1$, whereas Subplot B illustrates the optimal interval corresponding to the objective function $Z_2$ with $W_h=1, W_m=0$. In both cases, for a given value of $\alpha$ (outer color of the scatter points), we choose the parameter value $m_0$ (inner color of the scatter points) so that the annual mean prevalence is $\approx30\%$. The colorbars on the bottom illustrate the value of $m_0$ and $\alpha$ respectively. All other parameters are as in Table \ref{tab:white}.}}
    \label{fig:violin_ssn}
    \end{centering}
\end{figure}

The jump in optimal interval seen in Figure \ref{fig:S_MDA2_all}A, \ref{fig:gbl_sen}, and \ref{fig:violin} as we vary model parameters is related to the choice of the objective function. Regardless of the choice of model parameters (and hence disease prevalence), the effect of the drug on the blood-stage infected population ($I$) is to cause an instantaneous reduction at the time of the MDA corresponding to the effectiveness of the drug ($p_{blood}$). Hence, to minimise the blood-stage infected population alone, the optimisation will always suggest the immediate implementation of the second MDA round.  Similarly, the effect of the drug on the hypnozoite reservoir (which has an average size, $k_T$) is always to reduce its size regardless of model parameters (and disease prevalence). However, since the effect of the drug on $k_T$ will be more for larger hypnozoite reservoir sizes, to minimize the hypnozoite reservoir alone, the optimisation would suggest a longer interval (regardless of disease prevalence) so that the reservoir has time to build up before the next MDA round. In contrast, the effect of the drug on the liver-stage infected population ($L$) does vary with model parameters (and disease prevalence). For low disease prevalence, the average hypnozoite reservoir size will be smaller (see Equation (\ref{eqn:I_SS})) in which case $L$ will decrease at the time of the first MDA application. For higher disease prevalence, the average hypnozoite reservoir size will be larger and it is possible that $L$ will increase at the time of the first MDA application since those in $I$ have their blood-stage infection cleared but not all of their hypnozoites due to the large average hypnozoite reservoir size. The objective function, $Z_1$,  considers minimising both blood-stage infections ($I$) and hypnozoite burden within liver-stage infected fractions ($k_TL$). This increase in the liver-stage infected fractions when the first MDA is applied under higher disease prevalence means that a longer interval for the second MDA will be optimal to reduce the burden (Figure \ref{fig:S_MDA2_all}A, \ref{fig:gbl_sen}) while when prevalence is low the liver-stage infected fraction will decrease with the first MDA and a second MDA round within a short interval will be optimal. Furthermore, if we consider seasonality in the mosquito populations, the results are quite different. Figure \ref{fig:violin_ssn} illustrates the distribution of the optimal intervals ($x_0$ and $x_1$) when the annual mean prevalence is approximately $30\%$ for objective function $z_1$ (Figure \ref{fig:violin_ssn}A) and  for objective function $Z_2$ with $W_h=1, W_m=0$ (Figure \ref{fig:violin_ssn}B). The distribution of the optimal interval is consistent for both objective functions. The results indicate that when there are fluctuations in the abundance of mosquitoes in the environment, the optimal interval can be identified by measuring the prevalence.\par

\section{Discussion}\label{disscussion}

Targeting the hypnozoite reservoir is the most crucial action in any \textit{P. vivax} elimination strategy, as hypnozoites dominate \textit{P. vivax} transmission dynamics. In malaria elimination efforts around the world, interest in MDA using primaquine or tafenoquine has grown, as these are the only available drugs to treat liver-stage \textit{P. vivax} infections \cite{hsiang2013mass}. In this paper, we have developed a multiscale model that captures hypnozoite dynamics and the effect of the hypnozoite reservoir on disease transmission under radical cure treatment as a method of MDA. This model extends our previous work \cite{anwar2022multiscale} by integrating treatment into the model with multiple MDA rounds accounting for superinfection. We have extended Mehra \textit{et al.}’s \cite{mehra2022hypnozoite} within-host model by obtaining key parameters regarding hypnozoite dynamics under multiple MDA rounds and embedding these into a population-level transmission model that considers superinfection based on \cite{thesis_somya}. We have also included mosquito seasonality in our model to study the impact of MDA treatment when there is a seasonal effect on mosquitoes in the environment. According to our model and choice of parameters, MDA with radical cure can significantly reduce disease burden at the time the program is administered and maintain it at low levels when prevalence before the MDA intervention is low and if multiple MDA rounds are implemented (Figure \ref{fig:MDA6_sen}). Our model results are sensitive to some parameters, especially for parameter regimes where superinfection is likely. However, we found that the optimal MDA intervals for a specific objective depend on the parameter values (without seasonality), especially the ones that have more influence on the transmission dynamics (mosquitoes per human, hypnozoite activation rate, hypnozoite death rate, natural recovery rate, and average hypnozoite per mosquito bite). That is, even where different combinations of the model parameters correspond to the same equilibrium prevalence, the optimal intervals are not necessarily the same (without seasonality, Figure \ref{fig:violin}) and \ref{fig:gbl_sen}). However, when there is seasonal variation in the mosquito population in the environment, the optimal intervals are very similar (Figure \ref{fig:violin_ssn}) for different combinations of the model parameters that correspond to the same annual mean prevalence. Hence, prevalence alone should not be considered a reliable measure when determining optimal intervals between rounds of MDA, especially in regions where seasonal variation in the mosquito population is negligible.\par

Although the optimal interval, frequency, and population coverage with MDA are not clear in practice \cite{maude2012optimising,greenwood2008control,hsiang2013mass}, here we assume 100\% treatment coverage and that all drugs (both blood-stage and liver-stage) are 90\% effective. This assumption about the effectiveness of the radical cure drug is realistic, as studies show radical cure efficacy varies between 57·7\% and 95\% depending on the combination of drugs \cite{huber2021radical,nelwan2015randomized,llanos2014tafenoquine}. According to our model, the optimal intervals between MDA rounds vary with the prevalence before MDA, the number of MDA rounds under consideration, and the choice of the objective function (Figures \ref{fig:opt_MDA2_MDA3}, \ref{fig:S_MDA2_all}, \ref{fig:S_MDA3_all}). However, regardless of the objective and number of MDA rounds, the overall effect of the drug is only temporary under our model assumptions. This temporary effect is due to the assumption of the instantaneous effect of the drugs. This assumption is appropriate given that available drugs have half-lives varying from 3.7 hours to 28 days \cite{jittamala2015pharmacokinetic,SCHLAGENHAUF2019145} which is short compared to the time frame of interest (years).  Hence, in the long term, the dynamical system does not observe any drug effect and the system returns to its pre-MDA state, which is the expected outcome from a deterministic framework such as ours. A deterministic framework is useful to understand the disease dynamics for a large population size however for a small population size, it will be important to use a stochastic model to study disease-extinction scenarios \cite{allen2000comparison}. Currently, prophylaxis is not taken into account in our model. Accounting for prophylaxis might vary model outcomes, as a longer duration of prophylaxis leads to greater measured efficacy, especially in higher transmission settings \cite{huber2021radical}. Furthermore, given the mosquito population has a shorter lifespan, for a longer duration of prophylaxis a reasonable proportion of infectious mosquitoes may die out and disrupt the chains of transmission. The assumption of blood-stage infection clearance in the presence of superinfection is slightly different in the population model in comparison to the within host model. The within host model assumes that each blood-stage infection is cleared independently for analytical tractability \cite{mehra2022hypnozoite}. However, since we are not aware of any study that suggests that the blood-stage drugs act differently on each blood-stage infection, we assumed that the clearance of all blood-stage infections (regardless of how many there are) depends only on the efficacy of the drug, $p_{blood}$. \par

 Although being an effective intervention strategy, MDA has some disadvantages, especially in terms of drug resistance \cite{zuber2018multidrug,commons2018effect}. Because of the extensive use of antimalarial drugs, the parasite has developed resistance to some drugs, particularly chloroquine. However, chloroquine is still effective in most parts of the world for \textit{P. vivax} \cite{world20world}. Another challenge with MDA is the use of the anti-hypnozoicidal drugs primaquine and tafenoquine, as these can cause blood hemolysis in individuals with G6PD deficiency and problems in pregnant women \cite{howes2012g6pd,watson2018implications}. We do not consider G6PD deficiency in our model, but it could easily be extended to do so. We also do not consider drug resistance, immunity, or heterogeneity in bite exposure.\par

Since our model is deterministic, disease fade-out is not possible, but a disease in a real-life setting may undergo stochastically driven fade-out when the disease prevalence is sufficiently low \cite{keeling2011modeling,greenhalgh2015disease}. The primary purpose of this work is to optimise the implementation of the timing of the rounds of MDAs. However, disease elimination could be investigated with our multiscale model by approximating the elimination probability as a Binomial random variable. As \textit{P. vivax} parasites are transmitted through infectious mosquito bites, contributing to hypnozoites in the liver, it is as important to reduce mosquito-bite exposure or the abundance of mosquitoes as it is to clear hypnozoites from the liver \cite{le2007elaborated,price2020plasmodium,newby2015review}. Insecticide-treated nets, indoor residual spraying, and long-lasting insecticide–treated nets are some of the standard vector-control interventions for controlling malaria transmission and are necessary additional interventions alongside MDA as per the WHO guidelines \cite{zuber2018multidrug}. Including vector-control interventions with MDA and stochasticity in the model to obtain the probability of disease eradication is an avenue for potential future work.\par

To our knowledge, ours is the first multiscale model to provide a framework for studying the effect of multiple MDA rounds in both the within-host and population scale for \textit{P. vivax} transmission. The results from the model demonstrate the effect of several MDA rounds delivered at optimal intervals on both the transmission setting and hypnozoite dynamics. According to our model, \textit{P. vivax} transmission can only be interrupted for a certain period (the duration of which depends on the prevalence before MDA) when using MDA. That is, MDA alone is not sufficient to progress us towards sustained \textit{P. vivax} elimination under our model. While our model has not been parameterised for any particular geographical setting, it has the potential to aid policymakers in MDA control strategy decision-making. \par

\section{Funding}
M.N. Anwar is supported by a Melbourne Research Scholarship. J.M. McCaw’s research is supported by the Australian Research Council (DP170103076, DP210101920) and the NHMRC Australian Centre of Research Excellence in Malaria elimination (ACREME). J.A. Flegg’s research is supported by the Australian Research 
Council (DP200100747, FT210100034). \par

\section{Data availability}
Data sharing is not applicable to this article as no datasets were generated or analysed during the current study.

\appendix
\section{Model derivation with mosquito seasonality}\label{prop_model}
Let $X$, $Y$, and $Z$ represent the number of susceptible, blood-stage and liver-stage infected individuals and $U$, $V$, and $W$ represent the number of susceptible, exposed and infectious mosquitoes. Let $N_h=X+Y+Z$ be the total human population and $N_m(t)=U+V+W$ be the total mosquito population at time $t$, respectively. With mosquito seasonality, the model equations for the number of individuals in each compartment are:
\begin{align*}
    &\frac{\mathrm{d}X}{\mathrm{d}t}=-\lambda (t) X+\mu k_1(t)Z+p_1(t)\gamma Y+D_l(l)Z+p_1(t)D_b(t)Y,\\
     &\frac{\mathrm{d}Y}{\mathrm{d}t}=\lambda (t)(X+Z)+\alpha k_T(t)Z-\gamma (p_1(t)+ p_1(t)) Y-D_b(t)(p_1(t)+p_2(t))Y,\\
    &\frac{\mathrm{d}Z}{\mathrm{d}t}=-\lambda (t) Z-\mu k_1(t)Z-\alpha k_T(t)Z+p_2(t)\gamma Y-D_l(t)Z+(1-p(t))D_b(t)Z,\\
    &\frac{\mathrm{d}U}{\mathrm{d}t}=b_m(t)N_m(t)-ac\frac{Y}{N_h}U-gU,\\
    &\frac{\mathrm{d}V}{\mathrm{d}t}=ac\frac{Y}{N_h}U-(g+n)V, \\
    &\frac{\mathrm{d}W}{\mathrm{d}t}=nV-gW,
\end{align*}

where 
\begin{align*}
\lambda(t)=&\frac{N_m(t)}{N_h}ab\frac{W}{N_m(t)}=ab\frac{W}{N_h}.
\end{align*}
All model parameters are defined in Table \ref{tab:white}.
Here, $\mathrm{d}(N_h)/\mathrm{d}t=\mathrm{d}(X+Y+Z)/\mathrm{d}t=0$ so that the human population is constant in size over time. But for the mosquito population:
\begin{align*}
   \frac{\mathrm{d}N_m(t)}{\mathrm{d}t}=& \frac{\mathrm{d}(U+V+W)}{\mathrm{d}t},\\
   =&b_m(t)N_m(t)-g(U+V+W),\\
   =&b_m(t)N_m(t)-gN_m(t),\\
   =&b_m(0)\left(1+\eta \cos\left(\frac{2\pi t}{365}+\phi\right)\right)N_m(t)-gN_m(t),\\
   =&g\left(\eta \cos\left(\frac{2\pi t}{365}+\phi\right)\right)N_m(t),\\
  \therefore N_m(t)=&N_m(0)\text{exp}\left\{\frac{365g\eta}{2\pi} \sin\left(\frac{2\pi t}{365}+\phi\right)\right\},
\end{align*}
where $N_m(0)$ is the initial mosquito population size.

We now convert the above transmission model into a model on the proportion scale for consistency with the model given in Equations (\ref{eqn:pp2})-(\ref{eqn:pp6}).
Let $S=X/N_h,\ I=Y/N_h,\ L=Z/N_h,\ S_m=U/N_m(t),\ E_m=V/N_m(t),\ I_m=W/N_m(t)$. Therefore 

\begin{align*}
     &\lambda(t)=ab\frac{I_mN_m(t)}{N_h},\\
=&ab I_m\frac{N_m(0)}{N_h} \text{exp}\left\{\frac{365g\eta}{2\pi} \sin\left(\frac{2\pi t}{365}+\phi\right)\right\}\\
=&m_0ab I_m  \text{exp}\left\{\frac{365g\eta}{2\pi} \sin\left(\frac{2\pi t}{365}+\phi\right)\right\},\, [m_0=\frac{N_m(0)}{N_h}\text{ is the initial mosquito-human ratio}].
\end{align*}

\begin{figure}[ht]
    \begin{centering}
        \includegraphics[width=1\textwidth]{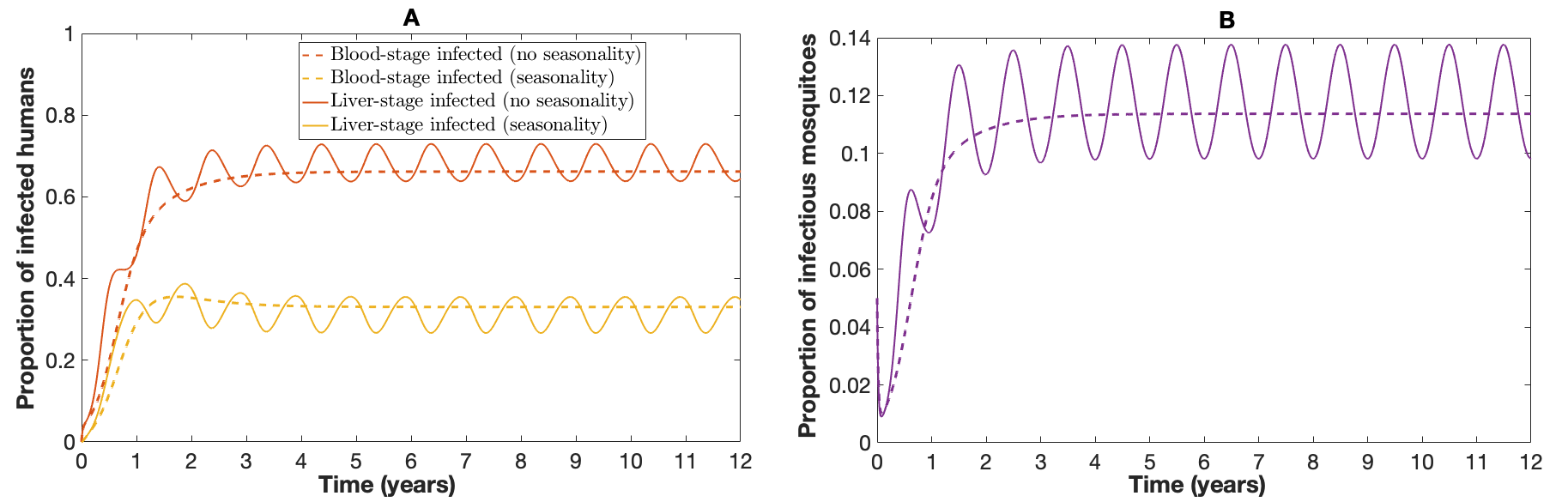}
    \caption{\textit{Model solutions with and without mosquito seasonality. Subplot (A) depicts the blood-stage and liver-stage proportions for humans over time with (solid lines) and without (dashed lines) seasonality. Subplot (B) illustrates the proportion of infectious mosquitoes with (solid lines) and without (dashed lines) seasonality over time. Parameters are as in Table \ref{tab:white}.}}
    \label{fig:prop_model}
    \end{centering}
\end{figure}

The equations for the human population on the proportion scale become:
\begin{align*}
    &\frac{\mathrm{d}(SN_h)}{\mathrm{d}t}=-\lambda(t) SN_h+\mu k_1(t)LN_h+p_1(t)\gamma IN_h+D_l(l)LN_h+p_1(t)D_b(t)IN_h,\nonumber\\
    \implies &\frac{\mathrm{d}S}{\mathrm{d}t}=-\lambda(t) S+\mu k_1(t)L+p_1(t)\gamma I+D_l(l)L+p_1(t)D_b(t)I.\\
    \text{Similarly,}\quad
     &\frac{\mathrm{d}I}{\mathrm{d}t}=\lambda(t)(S+I)+\alpha k_T(t)L-\gamma (p_1(t)+ p_1(t))I-D_b(t)(p_1(t)+ p_1(t))I,\\
    &\frac{\mathrm{d}L}{\mathrm{d}t}=-\lambda(t) L-\mu k_1(t)L-\alpha k_T(t)L+p_2(t)\gamma I-D_l(t)L+p_2(t)D_b(t)I.
\end{align*}
And the equations for the mosquitoes on the proportion scale become:
\begin{align*}
    &\frac{\mathrm{d}(S_mN_m(t))}{\mathrm{d}t}=b_m(t)N_m(t)-ac\frac{IN_n}{N_h}S_mN_m(t)-gS_mN_m(t),\nonumber\\
    \implies&N_m(t)\frac{\mathrm{d}S_m}{\mathrm{d}t}=b_m(t)N_m(t)-acIS_mN_m(t)-gS_mN_m(t)-S_m\frac{\mathrm{d}N_m(t)}{\mathrm{d}t},\nonumber\\
    \implies&\frac{\mathrm{d}S_m}{\mathrm{d}t}=b_m(t)-acIS_m-gS_m-\frac{S_m}{N_m(t)}(b_m(t)-g)N_m(t),\nonumber\\
   \implies &\frac{\mathrm{d}S_m}{\mathrm{d}t}=b_m(t)-acIS_m-b_m(t)S_m.\\
   \text{Similarly,}\quad 
    &\frac{\mathrm{d}E_m}{\mathrm{d}t}=acIS_m-\left(b_m(t)+n\right)E_m, \\
    &\frac{\mathrm{d}I_m}{\mathrm{d}t}=nE_m-b_m(t)I_m.
 \end{align*}
The model dynamics with mosquito seasonality in comparison with no seasonality are depicted in Figure \ref{fig:prop_model}. 

\section{Multiple infections given blood-stage infected}\label{MOI_I}
An individual might experience multiple blood-stage infections at the same time either due to bites from infectious mosquitoes or relapses from hypnozoite activation. We define the multiplicity of infection (MOI) as the number of distinct parasites co-circulating within a blood-stage infected individual. Thus, the multiplicity of infection (MOI) is given by the total number of bloods-stage infections (infections from mosquito bites and relapses) at time $t$: $M_I(t)=N_A(t)+N_P(t)$. Now, multiplicity of infection given empty hypnozoite reservoir: $M_I(t)|N_H(t)=0$ can be obtained from the PGF given by Equation (\ref{PGF_ntrt}) that holds for before treatment and by Equation (\ref{PGF}) which holds following treatment at times $s_1,\ s_2,\ \ldots, s_N$ as

\begin{align*}
  \mathbb{E}[z^{M_I(t)}|N_H(t)=0]=&\begin{cases}
    \frac{G(t,z_H=0,\ z_A=z,\ z_C=1,\ z_D=1,\ z_P=z\, z_{PC}=1)}{G(t,z_H=0,\ z_A=1,\ z_C=1,\ z_D=1,\ z_P=1\, z_{PC}=1)} & \text{if}\ t< s_1\\
    \frac{G^{s_1,s_2,\ldots s_N}(t,z_H=0,\ z_A=z,\ z_C=1,\ z_D=1,\ z_P=z\, z_{PC}=1)}{G^{s_1,s_2,\ldots s_N}(t,z_H=0,\ z_A=1,\ z_C=1,\ z_D=1,\ z_P=1\, z_{PC}=1)} & \text{if}\ t\ge s_N,\\
    \end{cases}\\
    =&\text{exp}\{h(z,t)-h(1,t)\},
\end{align*}

where
\begin{align}
   h(z,t)=&\begin{cases} 
   \int_0^t \lambda(\tau)\frac{ze^{-\gamma (t-\tau)}+(1-e^{-\gamma (t-\tau)})}{1+\big(p_H(t-\tau)+(1-z)p_A(t-\tau)\big)\nu }d\tau & \text{if}\ t< s_1\\
    \int_{s_{N}}^t \lambda(\tau)\frac{ze^{-\gamma (t-\tau)}+(1-e^{-\gamma (t-\tau)})}{1+\big(p_H(t-\tau)+(1-z)p_A(t-\tau)\big)\nu }d\tau\\+\int_0^{s_1} \lambda(\tau)\frac{z(1-p_{blood})e^{-\gamma (t-\tau)}+(1-(1-p_{blood})e^{-\gamma (t-\tau)})}{1+\big(p_H(t-\tau,s_1-\tau)+(1-z)p_A(t-\tau,s_1-\tau)\big)\nu }d\tau\\+\ldots+\int_{s_{N-1}}^{s_N} \lambda(\tau)\frac{z(1-p_{blood})^Ne^{-\gamma (t-\tau)}+(1-(1-p_{blood})^ne^{-\gamma (t-\tau)})}{1+\big(p_H(t-\tau,s_1-\tau,\ldots,s_n-\tau)+(1-z)p_A(t-\tau,s_1-\tau,\ldots,s_n-\tau)\big)\nu }d\tau
    & \text{if}\ t\ge s_N.
    \end{cases}\label{eqn:h}
\end{align}

Now, the probability mass function for $M_I(t)|N_H(t)=0$ is

\begin{align*}
&P(N_A(t)+N_P(t)=n|N_H(t)=0)=P(M_I(t)=n|N_H(t)=0)\nonumber\\
=&\text{exp}\left\{h(0,t)-h(1,t)\right\}\frac{1}{n!}\sum_{k=1}^n B_{n,k}\left[\frac{\partial h(0,t)}{\partial z},\frac{\partial^2 h(0,t)}{\partial z^2},\ldots,\frac{\partial^{n-k+1} h(0,t)}{\partial z^{n-k+1}}\right],
\end{align*}

where
\begin{align*}
   \frac{\partial^{k} h}{\partial z^k}(0,t)=&\begin{cases}
    k!\int_0^t \frac{\lambda(\tau)[\nu p_A(t-\tau)]^{k-1}}{\big[1+\nu \big(p_A(t-\tau)+p_H(t-\tau)\big)\big]^k}\Big(e^{-\gamma (t-\tau)}+\frac{\nu p_A(t-\tau)(1-e^{-\gamma (t-\tau)})}{1+\nu p_A(t-\tau)}\Big)d\tau& \text{if}\ t< s_1\\
    k!\Big(\int_{s_N}^t \frac{\lambda(\tau)\nu p_A(t-\tau)^{k-1}}{\big[1+\nu \big(p_A(t-\tau)+p_H(t-\tau)\big)\big]^k}\Big(e^{-\gamma (t-\tau)}+\frac{\nu p_A(t-\tau)(1-e^{-\gamma (t-\tau)})}{1+\nu p_A(t-\tau)}\Big)d\tau \\+\int_0^{s_1} \frac{\lambda(\tau)\nu {p_A^r(t-\tau,s_1-\tau)}^{k-1}}{\big[1+\nu \big(p_A^r(t-\tau,s_1-\tau)+p_H^r(t-\tau,s_1-\tau)\big)\big]^k}\Big((1-p_{blood})e^{-\gamma (t-\tau)}+\\\frac{\nu{p_A^r(t-\tau,s_1-\tau)}(1-(1-p_{blood})e^{-\gamma (t-\tau)})}{1+\nu {p_A^r(t-\tau,s_1-\tau)}}\Big)d\tau+\ldots\\+\int_{s_{N-1}}^{s_N} \frac{\lambda(\tau)\nu {p_A^r(t-\tau,s_1-\tau,\ldots,s_n-\tau)}^{k-1}}{\big[1+\nu \big(p_A^r(t-\tau,s_1-\tau,\ldots,s_n-\tau)+p_H^r(t-\tau,s_1-\tau,\ldots,s_n-\tau)\big]^k}\Big((1-p_{blood})^Ne^{-\gamma (t-\tau)}+\\\frac{\nu {p_A^r(t-\tau,s_1-\tau,\ldots,s_n-\tau)}(1-(1-p_{blood})^Ne^{-\gamma (t-\tau)})}{1+\nu {p_A^r(t-\tau,s_1-\tau,\ldots,s_n-\tau)}}\Big)d\tau\Big)
    & \text{if}\ t\ge s_n.\\
    \end{cases}
\end{align*}

\section{Steady state analysis (without seasonality)}\label{SS}

To obtain the steady state of the system without treatment and seasonality, first assume that the time-dependent parameters $p_1(t),\ p_2(t),\ k_1(t)$, and $k_T(t)$ are in steady state. We define
\begin{align*}
    \bar{p_1}=&\lim_{t\to\infty} p_1(t),\\
    \bar{p_2}=&\lim_{t\to\infty} p_2(t),\\
    \bar{k_1}=&\lim_{t\to\infty} k_1(t),\\
    \bar{k_T}=&\lim_{t\to\infty} k_T(t).\\
\end{align*}
Therefore at steady state, we have:

\begin{align}
    \label{eqn:SS1}
    &\frac{dS}{dt}=-\lambda S+\mu \bar{k_1}L+\bar{p_1}\gamma I=0,\\
    \label{eqn:SS2}
     &\frac{dI}{dt}=\lambda (S+L)+\alpha \bar{k_T}L-\gamma(\bar{p_1}+\bar{p_2}) I=0,\\
    \label{eqn:SS3}
    &\frac{dL}{dt}=-\lambda L-\mu \bar{k_1}L-\alpha \bar{k_T}L+\bar{p_2}\gamma I=0,\\
    \label{eqn:SS4}
    &\frac{dS_m}{dt}=g-acIS_m-gS_m=0,\\
    \label{eqn:SS5}
    &\frac{dE_m}{dt}=acIS_m-(g+n)E_m=0, \\
    \label{eqn:SS6}
    &\frac{dI_m}{dt}=nE_m-gI_m=0.
\end{align}

From Equations (\ref{eqn:SS3}), (\ref{eqn:SS4}) and (\ref{eqn:SS6}), at steady state we have:
\begin{align*}
L=&\frac{\gamma \bar{p_2}I}{\lambda+\mu \bar{k_1}+\alpha \bar{k_T}},\\
    S_m=&\frac{g}{g+acI},\\
    E_m=&\frac{gI_m}{n}.
\end{align*}
Substituting the value of $S_m$ and $E_m$ in Equation (\ref{eqn:SS5}), we get
\begin{align*}
    I_m=\frac{acnI}{(g+acI)(g+n)}.
\end{align*}
We have a constant human population, therefore
\begin{align*}
    S=&1-I-L\\
    =&1-I-\frac{\gamma \bar{p_2}I}{\lambda+\mu \bar{k_1}+\alpha \bar{k_T}}.
\end{align*}
Now, substituting the value of $S$ and $L$ in Equation (\ref{eqn:SS1}) gives:

\begin{align}
    -&\lambda \left(1-I-\frac{\gamma \bar{p_2}I}{\lambda+\mu \bar{k_1}+\alpha \bar{k_T}}\right)+\mu \bar{k_1} \frac{\gamma \bar{p_2} I}{\lambda+\mu \bar{k_1}+\alpha \bar{k_T}}+\bar{p_1}\gamma I=0,\nonumber\\
    \label{eqn:I_SS}
    \implies I=&\frac{\lambda(\lambda+\mu \bar{k_1}+\alpha \bar{k_T})}{(\lambda+\mu \bar{k_1})(\lambda+\bar{p_1}\gamma+\bar{p_2}\gamma)+(\lambda+\bar{p_1}\gamma)\alpha \bar{k_T}}.
\end{align}

To obtain a steady state prevalence, $I^*$, we need $\lambda=\lambda_{SS}$ to be at a steady state. Each of the parameters $\bar{p_1},\ \bar{p_2},\ \bar{k_1}$, and $\bar{k_T}$ can be expressed as a function of $\lambda_{SS}$ from the Equations (\ref{eqn:p_1_final}),\ (\ref{eqn:p_2}),\ (\ref{eqn:k1}), and (\ref{eqn:inf_sum}) respectively. That is,
\begin{align*}
    \bar{p_1}=&f_1(\lambda_{SS}),\\
    \bar{p_2}=&f_2(\lambda_{SS}),\\
    \bar{k_1}=&f_3(\lambda_{SS}),\\
    \bar{k_T}=&f_4(\lambda_{SS}),\\
\end{align*}
which makes Equation (\ref{eqn:I_SS}) a nonlinear function of $\lambda_{SS}$. That is

\begin{equation}    \label{eqn:ss_I}
      I^*=F(\lambda_{SS}).
\end{equation}

\begin{figure}
    \begin{centering}
        \includegraphics[width=\linewidth]{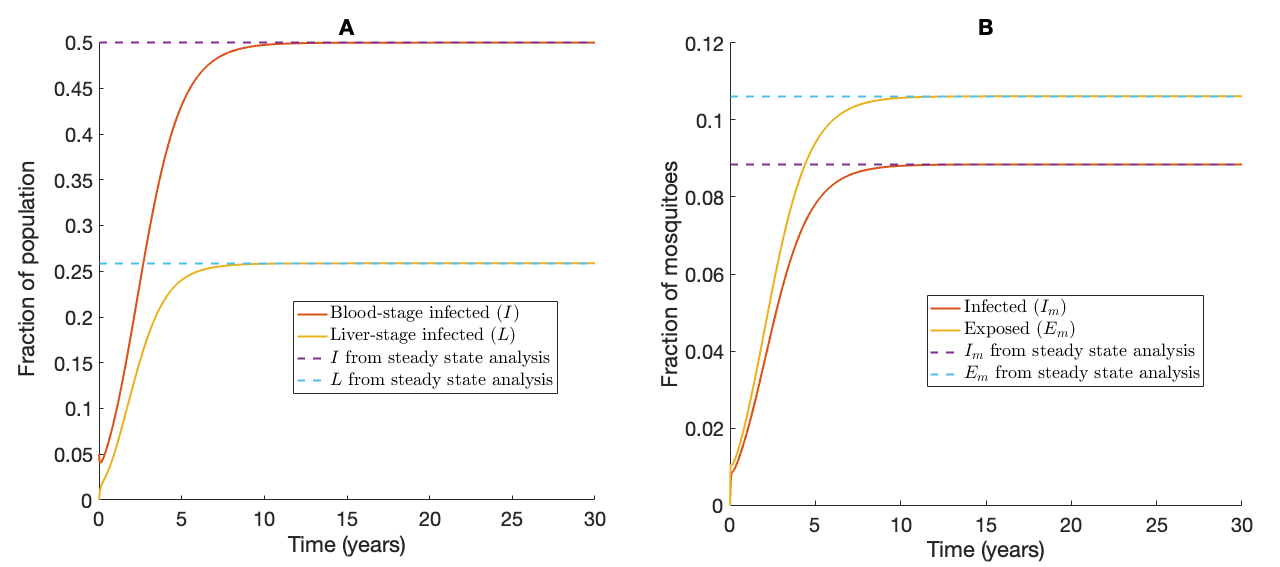}
    \caption{\textit{Temporal solution from numerical simulation of the model from time $t=0$ (solid lines) and steady state model beahviour from analysis (dash-dot lines) for  (A) the blood-stage and liver-stage human proportions and (B) the infected and exposed mosquito proportions. Parameters are as in Table \ref{tab:white}.}}
    \label{fig:SS_non_SS}
    \end{centering}
\end{figure}

Therefore, given a fixed $I^*$, we can solve Equation (\ref{eqn:ss_I}) to obtain $\lambda_{SS}$ and hence can find the other human and mosquito proportions at steady state. Note that the steady state disease prevalence, $I^*$, is obtained as a function of the human to mosquito ratio, $m$. That is, by varying $m$, we can vary $\lambda_{SS}$ and hence $I^*$. The steady state solution following this analysis is illustrated in Figure \ref{fig:SS_non_SS} and captures the long-term behaviour of the model based on numerical simulation.

\printbibliography[heading=bibintoc] 

\end{document}